\documentclass[journal,10pt,letterpaper,final,twocolumn]{IEEEtran}%
\usepackage{amsmath}
\usepackage{amsfonts}
\usepackage{cite}
\usepackage{amssymb}
\usepackage{graphicx}
\usepackage[usenames,dvipsnames]{color}
\usepackage{epstopdf}
\usepackage{lettrine}

\providecommand{\U}[1]{\protect\rule{.1in}{.1in}}
\pdfoutput=1
\providecommand{\U}[1]{\protect\rule{.1in}{.1in}}
\newtheorem{remark}{Remark}
\newtheorem{fact}{Fact}
\newtheorem{lemma}{Lemma}
\newtheorem{theorem}{Theorem}
\newtheorem{definition}{Definition}
\newtheorem{corollary}{Corollary}

\newcommand{\qed}{\nobreak \ifvmode \relax \else
      \ifdim\lastskip<1.5em \hskip-\lastskip
      \hskip1.5em plus0em minus0.5em \fi \nobreak
      \vrule height0.75em width0.5em depth0.25em\fi}
      
      \newcounter{myalgctr}
\newenvironment{claim}{
   \bigskip\noindent
   \refstepcounter{myalgctr}
 \newline  \textsc{{\bf{Claim} \themyalgctr:}}
   }{\newline  }  

\makeatother

\begin{document}

\title{Capacity and Power Scaling Laws for Finite Antenna MIMO Amplify-and-Forward  Relay Networks}

\author{\IEEEauthorblockN{David Simmons, \IEEEmembership{Student Member, IEEE,} Justin P. Coon, \IEEEmembership{Senior Member, IEEE,} and Naqueeb Warsi} \thanks{\IEEEauthorblockA{The authors are with the Department
of Engineering Science, University of Oxford, Parks Road, Oxford, OX1 3PJ, UK. 
 Email: david.simmons@eng.ox.ac.uk, justin.coon@eng.ox.ac.uk, naqueeb.ahmedwarsi@eng.ox.ac.uk}}\thanks{Copyright (c) 2014 IEEE. Personal use of this material is permitted.  However, permission to use this material for any other purposes must be obtained from the IEEE by sending a request to pubs-permissions@ieee.org.}}
\maketitle
\begin{abstract}
In this paper, we present a novel framework that can be used to study the capacity and power scaling properties of linear multiple-input multiple-output (MIMO) $d\times d$ antenna amplify-and-forward (AF) relay networks. In particular, we model these networks as  random dynamical systems (RDS) and calculate their $d$ Lyapunov exponents. Our analysis can be applied to systems with any per-hop channel fading distribution, although in this contribution we focus on Rayleigh fading. 
Our main results are twofold: 1) the total transmit power at the $n$th node will follow a deterministic trajectory through the network governed by the network's maximum Lyapunov exponent, 2) the capacity of the $i$th eigenchannel at the $n$th node will follow a deterministic trajectory through the network governed by the network's $i$th Lyapunov exponent.
Before concluding, we concentrate on some applications of our results. In particular, we show how the Lyapunov exponents are intimately related to the rate at which the eigenchannel capacities diverge from each other, and how this relates to the amplification strategy and number of antennas at each relay. We also use them to determine the extra cost in power associated with each extra multiplexed data stream. 
\end{abstract}

 {\section{Introduction}}


\lettrine{C}{onsider} a  multiple-input multiple-output (MIMO)  link with $d_S$ source antennas and $d_D$ destination antennas. It is well known that, under some basic assumptions (i.e., independent channel fading between each antenna pair), the capacity will \emph{almost surely} scale linearly with $\min\{  d_S , d_D \}$, \cite{telatar1999capacity}. 

\begin{figure}
\centering
\includegraphics[scale=0.38]{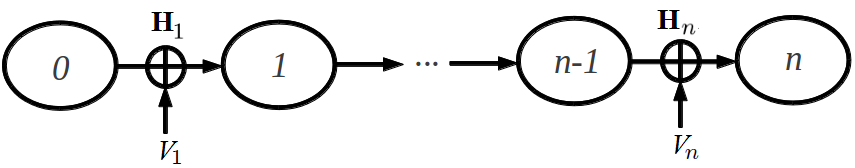}\caption{An illustration of a linear relay network. The $i$th hop's channel is described by the channel matrix $\mathbf H_i$. The  noise received at the $j$th node is described by the  vector $V_{j}$. Nodes $0$ and   $n$ are the source and destination, respectively.}
\label{fig:linearRelayNetwork}
\end{figure}

Now, consider an $n$-hop MIMO  link, aided by $n-1$ relay nodes\footnote{The deployment of relays is interesting because it can increase the diversity gain \cite{coopprotoutage}, and extend the coverage area of the network \cite{5733966}.}, where each relay node is equipped with $d$ transmit and receive antennas. Furthermore, assume that signals received at the $i$th relay node propagate only as far as the $(i+1)$th node (Fig. \ref{fig:linearRelayNetwork}). The end-to-end capacity, $c_n$, of such links has been studied in many works: see, e.g., \cite{5074422,6169203,fawaz2011asymptotic,DBLPGirnykVR14} for  amplify-and-forward (AF) studies, and \cite{5463228,4927440,xie2004achievable,levin2012amplify} for decode-and-forward (DF) studies. It is known, \cite{DBLP:journals/corr/KolteOG14}, that the capacity of such networks is achieved by employing DF relaying. However, when larger numbers of nodes are deployed, DF-based protocols may result in prohibitive latency/complexity because of the decoding process that takes place at each relay.  AF protocols become interesting at this point since they can be employed to yield low complexity and/or low latency solutions. Under certain scenarios, they also have the potential to offer greater diversity when compared to DF schemes~\cite{coopprotoutage}.  

The analysis of DF networks in the general multihop setting is made easier by the fact that a local view can often be taken -- i.e., the transmission is ``reset'' at each relay node, and thus sequential hops can, to a certain extent, be treated independently. This is not the case for AF networks, which must be observed \emph{globally} in the general case since the end-to-end transmission is affected by a composition of mappings, one for each hop.  Consequently, although AF relay networks may exhibit potential in multihop applications, relatively little is known about how these systems scale.

A number of results exist pertaining to  AF relay networks that  scale in size.
For a linear $n$-hop network, it was shown in  \cite{5074422,6169203}  that
 $\lim_{n\to\infty}$ $\left[\lim_{d_D\to\infty}\left[c_n/d_D\right]/n\right]$ exists \emph{almost surely} and is strictly positive, provided $d/d_D$ scales at  least linearly with $n$ (i.e., $d/d_D = \Omega\left(n\right)$). This work    also considered the aforementioned limit for other forwarding strategies; namely, DF,  compress-and-forward, and quantize-and-forward. In \cite{1013149}, the asymptotic (in matrix dimension) eigenvalue distribution of the channel's covariance matrix for linear $n$-hop MIMO channels with noiseless relays was established. Using free probability theory and, again, under the premise that negligible noise was received at the relays, it was shown in \cite{fawaz2011asymptotic} that when linear precoding was applied at each relay, $c_n$  would converge \emph{almost surely} to a  limit as $d$ grows large. The singular vectors of the optimal precoding matrices  for such a network when noise was negligible at the relays was also established in  \cite{fawaz2011asymptotic}. When noise was present at the relays, ergodic capacity and average bit error rate results were established in \cite{DBLPGirnykVR14} for multihop AF MIMO networks when arbitrary signaling occurs at the source node and, again, $d$ grows without bound.  Meanwhile,  in \cite{yeh2007asymptotic}, $c_n$ was assessed for general $n$-hop AF networks in terms of the limiting (in $d$) eigenvalue distribution of products of random matrices when noise was \emph{not} negligible at the relay nodes.  Related work on the diversity-multiplexing tradeoff, \cite{zheng2003diversity}, for various MIMO multihop relaying strategies can be found in \cite{4305387,5463228,yang2007diversity,sreeram2008dmt}. 
 
 Other attempts to determine the behavior of AF networks as they scale (not necessarily in the number of hops) can be found in \cite{4305403,6145524,1638664}. In more detail, \cite{4305403} considers a network of $m$ source-destination nodes communicating through a set of $k$ relays in a two-hop fashion. It is shown that, provided $k$ grows fast enough with $m$, in the large $m$ limit the network will ``crystallize'' into a set of nonfading  source-destination links with strictly positive capacity.  In \cite{6145524} a single-antenna source-destination pair aided by layered relays is studied. It is shown that such networks will approach the cut-set bound as the received power at each relay increases. An $m$ antenna source-destination pair assisted by $k$ single-antenna two-hop relays is studied in \cite{1638664}, where it is shown that for fixed $m$, the capacity of the network will obey $c=m/2\log k + O(1)$ as $k\to\infty$.

To the best of the authors' knowledge, all attempts to study the statistical behavior of the end-to-end capacity for $n$-hop ($n>2$) AF MIMO networks have leveraged a viewpoint in which the number of antennas at each node grows  large\footnote{For $n=2$ hops, results have been obtained for finite antenna systems (see, e.g., \cite{jin2010ergodic}).}. To achieve this,  results from random matrix theory \cite{tulino2004random} have commonly been employed; e.g.,  \cite{yin1986limiting,silverstein1995strong,marvcenko1967distribution}, which describe the asymptotic/limiting spectral properties of large random matrices. 
The limitations of such approaches are that statistical spectrum behavior is  established only at a macroscopic scale. The term macroscopic scale is commonly used in random matrix theory to describe macroscopic/`global' observables such as the empirical eigenvalue distribution (see \cite{guionnet2008large} for more details).
In this contribution, we establish statistical laws for each of the spectra individually. Crucially, this allows us to  determine capacity scaling laws for each of the subchannels of a network when the number of antennas at each node is \emph{finite}. This is done by employing the  formalism of  random dynamical systems\footnote{The RDS formalism can be applied to any relay system that 
can be described as a product of random matrices. This encompasses all AF strategies. Because of the `resetting' nature of digital relaying protocols (e.g., DF), it is unclear whether the RDS formalism can be employed to study the scaling properties of such networks.} (RDS),~\cite{arnold1998random}. Such systems have also been used to study econometrics \cite{bhattacharya2007random}, biological systems \cite{weinberger1982long,strogatz2001nonlinear}, chemical reactions \cite{strogatz2001nonlinear}, and the propagation of particles through fluidic media \cite{yu1990transition}. For relevant information on RDSs, the reader may refer to section \ref{sec:prelims}.

Going into more detail, the Lyapunov exponents~\cite{arnold1998random} of RDSs are known to characterize the exponential growth/decay rates of the spectrum of finite dimensional   random matrix products \cite{raghunathan1979proof,oseledets1968multiplicative,arnold1998random}. In this contribution, we use these Lyapunov exponents to study the spectral properties of the $n$-hop AF MIMO network\footnote{Lyapunov exponents were used in \cite{levy2009central,5605346}, where sum-capacity scaling laws were established for the non-ergodic Wyner cellular model as the number of cells grew large. In more detail, the upper Lyapunov exponent is used in \cite{levy2009central,5605346} by considering the Thouless formula, which relates the determinant of a large random matrix to a product of fixed size matrices.}.
The main result of our paper is that the Lyapunov exponents of the network, which are obtained by studying the network as an RDS, can be used to evaluate  the exponential growth/decay   of the $n$th node transmit power and $n$th node end-to-end eigenchannel capacity when each of the   nodes in the network has a finite number of antennas.

 {\subsection{Notation and Definitions\label{subsec:Notation}}}
We use $\mathbb {N,R}$ and $\mathbb{C}$ to denote the natural, real and complex numbers. We use $\overset{d}{=}$ to denote equality in distribution, $:=$ to denote equality by definition, logarithms are always to the base $e$ and $ \log^{+}(x) := \max\{0,\log x\}$.   $\bf{ 0}$ is used to denote the column vector of zeros, where the dimension of $\bf{ 0}$ will be implied from the context. Matrices are always represented using uppercase boldface notation,  vectors are always represented using uppercase non-boldface notation, and scalars are always represented using lowercase notation. $\mathcal E_i\{\mathbf{A}\}$ is used to denote the $i$th ordered eigenvalue of the matrix $\mathbf A$, where $\mathcal E_i(\mathbf A)\geq \mathcal E_j(\mathbf A)$ implies $i\leq j$. $\mathbf A^\dagger$ is used to denote the conjugate transpose of the matrix $\mathbf A$. Matrix products are defined in the following way:
\begin{equation}
\prod_{i = j}^{n}\mathbf{A}_i  :=  \mathbf{A}_n \cdots \mathbf{A}_j,
\end{equation}
and when $j=1$ we sometimes use the definition
\begin{equation}
\pi_n\left(  \mathbf A \right) :=  \prod_{i = 1}^{n}\mathbf{A}_i .
\end{equation}
The standard $2$-norm of a matrix $\bf A$ is denoted by $\left\Vert \bf A \right\Vert$, and its Frobenius norm is denoted by $\left\Vert \bf A  \right\Vert_F$.
The Landau notation $f(x)=o(g(x))$ is used to imply $\lim_{x\rightarrow\infty}f(x)/g(x) = 0$.
  Also, we use the following notation:
\begin{eqnarray}
\!f(n)\!\!\!\!\!&=&\!\!\!\!\!O\left(g(n)\right)\!\Rightarrow\!\exists k_1,\tilde{n}\!>\!0 \;\mathrm{s.t.}\; k_1|g(n)| > |f(n)|,\forall n>\tilde{n}\nonumber\\
\!f(n)\!\!\!\!\!&=&\!\!\!\!\!\Omega\left(g(n)\right)\!\Rightarrow\!\exists k_2,\tilde{n}\!>\!0 \;\mathrm{s.t.}\; k_2|g(n)| < |f(n)|,\forall n>\tilde{n}\nonumber\\
\!f(n)\!\!\!\!\! &=&\!\!\!\!\! \Theta\left(g(n)\right)\;\mathrm{if}\;f(n)=O\left(g(n)\right)\;\mathrm{and}\;f(n)=\Omega\left(g(n)\right);\nonumber
\end{eqnarray}
and, similar to the notation proposed in \cite{janson2011probability}, for a strictly positive random variable $f(n)$ depending on~$n$, and some $h(n)=|o(n)|$,
\begin{eqnarray}
\!f(n)\!\!\!\!\!&=&\!\!\!\!\!O_{\mathbb P}\left(g(n)\right)\;\;\Rightarrow  \; \lim_{n\to\infty}\!\! \mathbb P\! \left[    f(n)   \leq\!    g(n) e^{h(n)}   \right] = 1\nonumber .\nonumber\\
\!f(n)\!\!\!\!\!&=&\!\!\!\!\!\Omega_{\mathbb P}\left(g(n)\right) \;\;\Rightarrow   \;\lim_{n\to\infty}\!\! \mathbb P\! \left[    f(n)   \geq \!    g(n) e^{-h(n)}   \right] = 1 \nonumber\\
\!f(n)\!\!\!\!\! &=&\!\!\!\!\! \Theta_{\mathbb P}\left( g(n)\right)\;\mathrm{if}\;f(n)=O_{\mathbb P}\left(g(n)\right)\;\mathrm{and}\;f(n)=\Omega_{\mathbb P}\left(g(n)\right).\nonumber
\end{eqnarray}

 {\subsection{Paper Layout\label{subsec:layout}}}
Section \ref{sec:System-Model} introduces the system model. Section \ref{sec:keyresults} clarifies the key results obtained in this paper. Section \ref{sec:prelims} introduces the mathematical preliminaries and new RDS results that will be utilized throughout this paper. In section  \ref{sec:PwrCap} we calculate the Lyapunov exponents, and show that they can be used to characterize the network's transmit power and end-to-end eigenchannel capacity. Section   \ref{sec:ECapDiv} establishes applications of the results that are obtained in section \ref{sec:PwrCap}. Section \ref{sec:Illus} provides numerical illustrations of the theory that has been developed.  Finally,  section \ref{sec:conc} concludes the paper.

 {\section{System Model  \label{sec:System-Model}}}

Let us present the signaling model used in this paper. 
Consider an $n$-hop AF relay network, as depicted in Fig.~\ref{fig:linearRelayNetwork}.
 We assume that each node has $d\geq 1$ transmit and receive antennas. Independent frequency-flat Rayleigh fading \cite{goldsmith2005wireless} is assumed to take place between each node pair. Thus, the channel for the $i$th hop can be described by a $d\times d$ random matrix, $\mathbf H_{i}$, whose elements are  circularly symmetric complex Gaussian   \cite{goldsmith2005wireless} with total variance  $\mu_i$; i.e., the $(a,b)$th element of $\mathbf H_{i}$ is given by  $h_{ab,i}\sim\mathcal{CN}(0,\mu_i)$.

At each node (apart from the zeroth node; i.e., the source) we assume noise is introduced into the system. We use $ {V}_j\in \mathbb C^d$ to denote the vector of noise terms introduced at the $j$th relay. The elements of $ {V}_{j}$ correspond to the noise samples received at each antenna of node $j$ and are independent complex Gaussian random variables with total variance $n_0$. 

An information vector 
\begin{equation}
X_0 = \left[ x_{0,1}, \dots , x_{0,d}\right]^T\label{eq:SourceVector}
\end{equation}
 is constructed at the source (node $0$). We assume each element of $X_0$  has a mean of zero and average power given by $\mathbb E[|x_{0,i}|^2]=p_0/d$. The $i$th element of $ {X}_0$ is then transmitted from the $i$th antenna of node $0$.

We assume the $j$th relay node receives the transmission only from the $(j-1)$th node in one time slot. This relay then applies a scalar gain, $\alpha_j$, to the received signal on each of its antennas and transmits in the next time slot. Thus, the relays operate in a half-duplex manner.
The gain for the $j$th relay is either a fixed-gain parameter, depending only upon the average statistics of the channel matrix of the previous hop, given by
\begin{equation}
f_{ j}  = \sqrt{\frac{  { p}_j}{p_{j-1} d \mu_j +d n_0}};\label{eq:FG}
\end{equation}
or a variable-gain parameter given by \cite[eq. (7)]{5089994}
\begin{equation}
v_{ j}  = \sqrt{\frac{   {p}_j}{\frac{p_{j-1}}{d}\left\Vert\mathbf H_j  \right\Vert_F^2 + dn_0}}.\label{eq:VG}
\end{equation}
The term $p_{j}$ is selected by the relay, and  represents the average transmit power at  node $j$. Also, we assume that $\lim_{n\to\infty} ({1}/{n})\log( {p_n}/{p_0}) < \infty$. This assumption implies that the average transmit power does not grow at a super-exponential rate. Similar assumptions have also been made in \cite{6169203}.
It is important to note that we are implicitly assuming the relays have access to statistical channel state information in the form of $\mu_i$ for fixed-gain and $\left\Vert\mathbf H_j  \right\Vert_F^2$ for variable-gain. The precise mechanism by which these are obtained is beyond the scope of this paper. Needless to say, using tools such as received signal strength indicators, it is possible for these to be gleaned from a channel output without having to perform decoding operations at each relay. 

The information bearing content of the signal (herein referred to as the \emph{information component}) at the $n$th node is given by
\begin{align}
\label{eq:SISO_Infor_Form}
\mathcal{I}_{n}^{(\alpha)} &=  \alpha_n\mathbf{H}_n\mathcal{I}_{n-1}^{(\alpha)}\\
& =  \prod_{j=1}^{n}\alpha_{j}\mathbf H_{j}X_0,
\end{align}
where $\alpha  \in\{f,v\}$ dependent upon whether fixed-gain or variable-gain is being implemented.
Similarly, the total transmitted signal at the $n$th node is given by
\begin{align}
X_{n}^{(\alpha)}&=  \alpha_n \left(        \mathbf{H}_n X_{n-1}^{(\alpha)} + V_n        \right) \\
&=\mathcal{I}_{n}^{(\alpha)} + \underset{\mathcal{N}_n^{(\alpha)} }{\underbrace{\alpha_n \sum_{i=1}^{n}\prod_{j=i+1}^{n}\alpha_{j-1} \mathbf H_{j} V_{i}}}  ,
\label{eq:x}
\end{align}
 where  $\prod_{j=n+1}^{n}\alpha_{j-1} \mathbf H_{j}  V_n :=    V_n$ and  $\mathcal{N}_n^{(\alpha)}$ denotes the accumulated noise at node $n$. 
  Owing to our choice of gain, and by the definition of the source transmit vector \eqref{eq:SourceVector}, for all $i\in\mathbb{N}\cup \{0\}$ the $i$th node is subject to the following average power constraint:
 \begin{equation}
 \mathbb E \left[  X_i^{(\alpha)\dagger}  X_i^{(\alpha)}  \right]   \leq   p_i.
 \end{equation} 

The channel input,  \eqref{eq:x}, can be re-expressed as the first $d$ entries of the following matrix product:
\begin{equation}
\left[ \begin{array}{c}
X_n^{(\alpha)}\\
\sqrt{n_0}\\
\end{array} \right]=\prod_{j=1}^{n}\mathbf Q_j^{(\alpha)}\left[ \begin{array}{c}
X_0\\
\sqrt{n_0}\\
\end{array} \right],
\label{eq:Mx}
\end{equation}
where
\begin{equation}
\mathbf Q_j^{(\alpha)}:= \left[ \begin{array}{cc}
\alpha_{j} \mathbf H_{j} & \alpha_{j} Z_{j}\\
\bf{0}^T & 1
\label{eq:M}
\end{array} \right]
\end{equation}
  and $Z_i = V_i / \sqrt{n_0}$. This matrix formulation will help us to establish power scaling laws for the network.
  
Finally, we give a definition for the capacity of the network described above.
\begin{definition}\label{def:capacity_nhop}
The information capacity of the $n$-hop network with amplification strategy $\alpha$ and average power constraint $p_i$ for the $i$th node is
\begin{equation}
c_n^{(\alpha)} = \max_{P_{X_0}\left(  X \right) } I \left(  X_0 ; X_n^{(\alpha)} /\alpha_n  \right),\label{eq:MutualInfoCap}
\end{equation}
where $P_{X_0}$ denotes the probability density function of the source vector, $X_0$, and $I\left(X_0; X_n^{(\alpha)}/\alpha_n \right)$ denotes the mutual information between $X_0$ and $ X_n^{(\alpha)} / \alpha_n$.
\end{definition}
\begin{fact}\label{fact:capacity_nhop}
 From \cite{6169203}, the end-to-end channel capacity  of the network, \eqref{eq:MutualInfoCap}, is given by
  \begin{align}
c_n^{(\alpha)} & =  \log\det\left(  \mathbf{I}_d + \mathbf R_{\mathcal{I},n}^{(\alpha)}\mathbf R_{\mathcal{N},n}^{(\alpha)-1}\right) 
 \\&=\sum_{i=1}^{d}c_{n,i}^{(\alpha)}\;\mathrm{(nats / channel\; use)} , \label{eq:C}
\end{align}
where
\begin{align}
\mathbf R_{\mathcal{I},n}^{(\alpha)} &= \left(p_0\prod_{i=1}^{n}{\alpha_{i}^2}\right) \mathbf H_{n}\cdots\mathbf H_{1}\mathbf H_1^\dagger \cdots\mathbf H_{n}^\dagger ,\label{eq:RInalpha}\\
\mathbf R_{\mathcal{N},n}^{(\alpha)} &= n_0 \left( \mathbf{I}_d + \sum_{l=2}^{n} \prod_{i=l}^{n}\alpha_{i}^{2} \mathbf H_n\cdots \mathbf H_l \mathbf H_l^\dagger\cdots\mathbf H_n^\dagger  \right)\;\;\label{eq:RNnalpha}
\end{align}
are covariance matrices; and 
$
c_{n,i}^{(\alpha)}=\log\left( 1 + \mathcal E_i\left(\mathbf R_{\mathcal{I},n}^{(\alpha)} \mathbf R_{\mathcal{N},n}^{(\alpha)-1} \right) \right) \label{eq:Ecap}
$
is the capacity of the $i$th eigenchannel at the $n$th node. 
\end{fact}

 {\section{Key Results\label{sec:keyresults}}}

One of the key insights that we provide in this paper is that $n$-hop AF MIMO systems can be studied from the viewpoint of RDS. To the best of our knowledge, this is the first time such an approach has been taken in the literature. This viewpoint then leads us to obtain the following results:

\begin{itemize}
\item In Lemma \ref{lem:AVGPWR} and Theorem \ref{lem2}, we show that the $d\times d$ antenna MIMO AF network has associated with it an ordered set $\{\lambda_{\alpha\mathbf{H},1} ,\cdots,\lambda_{\alpha\mathbf{H},d} \}$ of Lyapunov exponents satisfying
$$\lambda_{\alpha\mathbf{H},1} >\cdots>\lambda_{\alpha\mathbf{H},d} ,$$
where the   $\alpha\in\{f,v\}$ term in the subscript denotes the amplification strategy that is being implemented (i.e., $f$ for fixed-gain or $v$ for variable-gain).
From this ordered set, two other sets of exponents are established. The first of these sets is constructed from elements of the form $\lambda_{\mathbf{Q},i}^{(\alpha)} = \max\{\lambda_{\alpha\mathbf{H},i},0\}$, and is associated with the instantaneous total transmit signal at the $n$th node. The second of these sets is constructed from elements of the form  $\lambda_{\gamma,i}^{(\alpha)} = \min\{2\lambda_{\alpha\mathbf{H},i} ,0\}$, and is associated with the end-to-end SNR of the network's $d$ eigenchannels.
\item In Lemma \ref{lem:AVGPWR}, we show that the instantaneous transmit power, $\Vert X_n^{(\alpha)}  \Vert^2$, at the $n$th node obeys the relationship
\begin{equation}
\Vert X_n^{(\alpha)}  \Vert^2 = \Theta_\mathbb{P}\left(  e^{2n\lambda_{\mathbf{Q},1}^{(\alpha)}}  \right),\label{eq:Xtrajectory}
\end{equation}
where,  $f(n) = \Theta_{\mathbb P}(g(n))$ implies that $f(n)$ is equal to $g(n)$, to first order in the exponent \cite[eq. (3.26)]{cover2012elements}. This is defined more rigorously in the notation subsection (subsection \ref{subsec:Notation}).
\item In  Theorem \ref{lem2}, we show that the SNR and capacity of the $i$th eigenchannel at the $n$th node, $\mathcal E_{i,n}\left(  \mathbf{R}_{\mathcal I , n}^{(\alpha)}\mathbf{R}_{\mathcal N,n}^{{(\alpha)}-1}  \right) $ and $c_{i,n}^{(\alpha)} $,  satisfy the relations
\begin{align}
\mathcal E_{i,n}\left(  \mathbf{R}_{\mathcal I , n}^{(\alpha)}\mathbf{R}_{\mathcal N,n}^{{(\alpha)}-1}  \right) &= \Theta_{\mathbb P}\left(  e^{n\lambda_{\gamma,i}^{(\alpha)}}  \right)\;\nonumber\\\mathrm{and}\qquad c_{i,n}^{(\alpha)}  &= \Theta_{\mathbb P}\left(  e^{n\lambda_{\gamma,i}^{(\alpha)}}  \right),\label{eq:Ctrajectory}
\end{align}
where $\mathbf{R}_{\mathcal I , n}^{(\alpha)}$ and $\mathbf{R}_{\mathcal N,n}^{{(\alpha)} }$ are given in \eqref{eq:RInalpha} and \eqref{eq:RNnalpha}, respectively.
\end{itemize}
On top of our main results, we also establish the following notable secondary results:
\begin{itemize}
\item In Lemma \ref{lem:AVGPWR}, we show that to ensure    the instantaneous transmit power \emph{almost surely} displays no exponential growth, and that the end-to-end capacity of the dominant eigenchannel \emph{almost surely} displays no exponential decay (i.e., from \eqref{eq:Xtrajectory} and \eqref{eq:Ctrajectory}, $\lambda_{\alpha\mathbf{H},1}  = 0$), the \emph{average} transmit power \emph{must} grow exponentially with $n$. Furthermore, this rate of average power growth can be reduced by:
\begin{enumerate}
\item   implementing variable-gain instead of fixed-gain, 
\item increasing the number of antennas at each node.
\end{enumerate}
\item  In \eqref{eq:CnumLya}, we show that the exponential rate at which the capacities of the $i$th and $j$th ($i<j$) eigenchannels diverge away from each other is given by $n\left(\lambda_{\gamma,i}^{(\alpha)} -\lambda_{\gamma,j}^{(\alpha)} \right)$. When the $i$th and $j$th eigenchannel capacities are either both decaying or both not decaying, this divergence rate is shown to be independent of whether fixed-gain or variable-gain relaying is being performed. Furthermore, from Lemma \ref{lem:1stOrder} and Corollary \ref{cor:9},  with $i=1$, to ensure that this rate is asymptotically bounded away from infinity (so that multiplexing $j$ streams is asymptotically viable) we must either: 
\begin{enumerate}
\item ensure that $\lambda_{\alpha\mathbf{H},j} \geq0$, 
\item ensure that the number of antennas at each node grows  like $d=\Omega (n)$. This result complements those presented in \cite{5074422,6169203}.
\end{enumerate}
\item In Remark \ref{rem:cost}, we assign a transmit power cost to the $n$th node for each extra data stream that is multiplexed over the network. In particular, if $i$ data streams are being multiplexed, then, to multiplex one extra stream, we must increase the $n$th relay's instantaneous transmit power by  a factor of $\exp(n/(d-i))$.
\end{itemize}

On the way to proving the above mentioned results, we also obtain the following    RDS results, which we believe are of independent interest.

\begin{itemize}
\item For $i\in\mathbb N$, let   $\mathbf A_i\in\mathbb C^{d\times d}$ and $R_i\in\mathbb C^d$  be random matrices and vectors, respectively, with $\mathbb E\log^+\Vert\mathbf A_1\Vert<\infty$, $\mathbb E\log^+\Vert R_1\Vert<\infty$, and $R_i\overset{a.s.}{\neq} 0 $. Suppose that there exists $\alpha_j,\beta_j\in\mathbb R$ such that $\mathbf A_1$ is equal in distribution to $\alpha_j \mathbf A_j$ and $R_1$ is equal in distribution to $\beta_j R_j$. In Lemma \ref{lem:AlwaysPos}, we show that the Lyapunov exponents of an affine RDS taking the form 
 \begin{equation}
X_n = \mathbf A_n X_{n-1} + R_n, \label{eq:affinekeyres}
\end{equation}  
  are strictly positive, and, consequently, are  identical to those of
   \begin{equation}\!\!\!\!\!
 \left[\begin{array}{cc}
  X_n \\
 1
\end{array}\right] =  \left[\begin{array}{cc}
\mathbf A_n & R_n\\
\bf{0}^{T} & 1
\end{array}\right] \cdots  \left[\begin{array}{cc}
\mathbf A_1 & R_1\\
\bf{0}^{T} & 1
\end{array}\right] \left[\begin{array}{cc}
  X_{0} \\
 1
\end{array}\right].\label{eq:nonaffinekeyres}
\end{equation}
\item  In Theorem \ref{lem:TTP}, we show that the Lyapunov exponents of  \eqref{eq:nonaffinekeyres} are given by the non-negative Lyapunov exponents of 
\begin{equation}
\pi_n(\mathbf{A}) :=\mathbf A_{n}\cdots\mathbf A_{1} .
\end{equation}
\end{itemize}

Less formally, our RDS results  provide us with a framework for determining \emph{all} of the Lyapunov exponents of $d$ dimensional affine RDSs, which will be crucial to our information theoretic analysis.

 {\section{Random Dynamical Systems}\label{sec:prelims}}
In this section, we introduce the RDS results that will be relied upon heavily throughout this paper. The first subsection is devoted to presenting preexisting RDS theory, while the second subsection presents a new result which will be used to calculate the Lyapunov exponents of affine systems.

 {\subsection{Preliminary RDS Results}}

The study of dynamical systems is concerned with tracking the trajectory
of a position (particle/state/point) through a state space. In the discrete case, this position
is calculated through the repeated action of a deterministic map. 
Informally, an RDS occurs when this map is non-deterministic
and drawn from a sample space according to some fixed probability distribution. Such systems are are often used to study econometrics \cite{bhattacharya2007random}, biological systems \cite{weinberger1982long,strogatz2001nonlinear}, chemical reactions \cite{strogatz2001nonlinear}, and the propagation of particles through fluidic media \cite{yu1990transition}. 
The formal and rather intricate definition of an RDS can be found in~\cite{arnold1998random}.

In this contribution, we consider an RDS to be the action of a product of $d\times d$ $(d\in\mathbb N)$ complex random matrices on an appropriately dimensioned vector (the initial state $X_0\in \mathbb C^d$). The state of the RDS  at time $n$ ($X_n\in \mathbb C^d$) can then be written as either
 \begin{equation}
 X_n =\mathbf A_{n}\cdots \mathbf A_{1} X_0,\label{eq:forwardRDS}
 \end{equation}
 or 
  \begin{equation}
 X_n =\mathbf A_{1}\cdots \mathbf A_{n} X_0\label{eq:backwardRDS}
 \end{equation}
where, in general, we assume that $\mathbf A_1,\ldots,\mathbf  A_n$ are independent and identically distributed (i.i.d.) up to an arbitrary positive scaling factor. Mathematically, this means that  $\exists \;\alpha_{i}>0$ such that $\mathbf A_1 \overset{d}{=} \alpha_i \mathbf A_i$ for all $i$. Eqs. \eqref{eq:forwardRDS} and \eqref{eq:backwardRDS} are referred to as  forward and backward  RDSs, respectively, and take  their names from the  {forward}~\cite[Def'n. 1.1.1]{arnold1998random} and  {backward}~\cite[Rem. 1.1.10]{arnold1998random}  cocycle properties that their random mappings satisfy. It is interesting and important to note that, unlike \eqref{eq:forwardRDS}, \eqref{eq:backwardRDS} is somewhat unnatural, in the sense that it is anticausal; however, all of the RDS properties that are to be described for \eqref{eq:forwardRDS} will apply to \eqref{eq:backwardRDS} as well, \cite{arnold1998random}.

Suppose we wish to study the asymptotic behavior of
\begin{equation}\label{eq:pi}
\left\Vert  \pi_n(\mathbf A) \right\Vert 
\end{equation}
as $n\rightarrow\infty$. A traditional approach is to exponentiate the logarithm of the norm; i.e., write \eqref{eq:pi} as
\begin{equation}
  \left\Vert  \pi_n(\mathbf A) \right\Vert = e^{n\frac 1 n \log \left\Vert  \pi_n(\mathbf A) \right\Vert }
\end{equation}
and investigate the behavior of the exponent, specifically, the term $\frac{ 1}{ n} \log \left\Vert  \pi_n(\mathbf A) \right\Vert $ as $n$ grows large.  In this manner, the exponential growth/decay rate of the system can be observed.  If $\{ \mathbf A_{j}\} $ was a set of scalars (i.e., $d=1$), the law of large numbers could be employed to evaluate the limiting behavior of $\frac{ 1}{ n} \log \left\Vert  \pi_n(\mathbf A) \right\Vert $; however, this is not the case for general $d$.

The question of whether $\frac{1}{n}\log\left\Vert  \pi_n(\mathbf A) \right\Vert  $
tends to a limit does not have a clear answer in most cases. Under the condition that $\mathbb{E}[\log^{+}\left\Vert \mathbf A_{1}\right\Vert ]<\infty$ and $ \lim_{n\to\infty}1/n\sum_{i=1}^{n}\log^+ |\alpha_{i}|<\infty$, however, the theorem of Furstenberg and Kesten \cite{arnold1998random}
guarantees that the limit exists. We then obtain the \emph{Lyapunov index}:
\begin{definition}The Lyapunov index is given by
\begin{equation}\label{eq:Lyapunov_index_def}
\iota(\mathbf A):=\limsup_{n\to\infty} \frac{1}{n}\log\left\Vert  \pi_n(\mathbf A) \right\Vert .
\end{equation}
\end{definition}

The Lyapunov index can be used
to describe the exponential growth rate of $\left\Vert\pi_n(\mathbf A)\right\Vert$.  By evaluating the Lyapunov index at
a specific initial position within the state space, we then obtain the \emph{Lyapunov exponent}:

\begin{definition}\label{def:LE}The Lyapunov exponent is given by
\begin{equation}\label{eq:Lyapunov_exp_def}
\lambda(\mathbf A,X):=\limsup_{n\to\infty} \frac{1}{n}\log\left\Vert \pi_n(\mathbf A) X\right\Vert.
\end{equation}
\end{definition}

The Lyapunov exponent can be used
to describe the exponential growth rate of the norm of a trajectory through its state space, where the initial state of the trajectory is given by $X$.

\begin{remark}
In the definitions of the Lyapunov index and exponent (\eqref{eq:Lyapunov_index_def} and \eqref{eq:Lyapunov_exp_def}, respectively), if the system is linear then the  $\lim\sup$ can be replaced with a limit, \cite{arnold1998random}.
\end{remark}

\begin{fact}\label{lem:props}
From \cite[pp. $114-115$, Theorem 3.3.3]{arnold1998random}, assuming   $\mathbb{E}[\log^{+}\left\Vert \mathbf A_{1}\right\Vert ]<\infty$ and $-\infty < \lim_{n\to\infty}1/n\sum_{i=1}^n\log^+|\alpha_i| <\infty$ (where $\mathbf{A}_1\overset{d}{=}\alpha_i\mathbf{A}_i$), the Lyapunov exponent has
the following properties~\cite{arnold1998random}:
\begin{enumerate}
\item $\lambda(\mathbf A,X)\in\mathbb{R}\cup\{ -\infty\} $
$\forall\; X\in\mathbb{C}^{d}$, where   $\lambda(\mathbf A,\bf{0}):=-\infty$;
\item \label{enu:Number_of_Lyapunov_Exponents}The number, $p$, of distinct
values, $\lambda_i$, that $\lambda(\mathbf A,X)$ can take on for $X \in \mathbb{C}^{d}\!\setminus\!\{\bf{0}\}$ is at most
$d$, and we have $-\infty\leq\lambda_{p}<\cdots<\lambda_{1}<\infty$.
\item \label{filtration}The sets 
\begin{equation}
\mathcal{V}_i := \left\{  X: \lambda\left( \mathbf A , X\right) \leq\lambda_i\right\} 
\end{equation}
are linear subspaces, form a filtration
\begin{equation}
\{0\}=:\mathcal{V}_{p+1} \subset \mathcal{V}_p\subset\cdots\subset \mathcal{V}_1 = \mathbb C^d
\end{equation}
(where all inclusions are proper), and 
\begin{equation}
\lambda_{i} = \lambda\left( \mathbf A,X\right)\Leftrightarrow X\in \mathcal{V}_i\setminus \mathcal{V}_{i+1},\;i=1,\dots,p.
\end{equation}
The integer $m(i) := \dim \mathcal{V}_i - \dim \mathcal{V}_{i-1}$ is the multiplicity of $\lambda_i$.
\item \label{itemSV}The limiting behavior for the ordered singular values of the matrix product $\pi_n(\mathbf A)$ satisfies
\begin{equation}
\frac{1}{2n}\log \mathcal E_i\left\{ \pi_n(\mathbf A)\pi_n(\mathbf A)^\dagger\right\}\to \lambda_i.\label{item3}
\end{equation}
Consequently,
the random variable  $\mathcal E_i\left\{ \pi_n(\mathbf A)\pi_n(\mathbf A)^\dagger\right\}$ satisfies
\begin{equation}
\mathcal E_i\left\{ \pi_n(\mathbf A)\pi_n(\mathbf A)^\dagger\right\} = \Theta_{\mathbb P} \left(  e^{2n\lambda_i}\right).\label{lem:traj}
\end{equation}
\end{enumerate}
  \end{fact}
  
 In what follows, we will often drop the functional notation $\lambda(\mathbf A,X)$ and simply write $\lambda_{\mathbf A,i}$ to refer to the $i$th ordered Lyapunov exponent of the system corresponding to $\pi_n\left(\mathbf A\right)$. When it is clear, we may also omit the subscript $\mathbf A$ as we did in Fact \ref{lem:props}.

 {\subsection{On the Lyapunov Exponents of Affine RDS\label{sec:afflin}}}



Throughout this paper, we will often be concerned with the Lyapunov exponents of affine systems of the form
 \begin{equation}
X_n = \mathbf A_n X_{n-1} + R_n, \label{eq:affine}
\end{equation}
where $\mathbf A_n\in\mathbb C^{d\times d}$ is a random matrix and $R_n\in\mathbb C^d$ is a random vector that satisfies $R_n\overset{d}{=}-R_n$  and $R_n\overset{a.s.}{\neq} 0$.  
The following theorem will be used in the calculation of these exponents. 
 \begin{theorem} \label{lem:TTP}
For $i\in\mathbb N$,  consider the product of random matrices $\pi_n(\mathbf M)$, where
\begin{equation}
\mathbf M_i =  \left[\begin{array}{cc}
\mathbf A_i & R_i\\
\bf{0}^{T} & a_i
\end{array}\right]\in\mathbb C^{(d+1)\times(d+1)} 
\end{equation}
  $\mathbb E\log^+\left\Vert\mathbf A_i  \right\Vert <\infty$, $\mathbb E\log^+\left\Vert  R_i  \right\Vert <\infty$, and $\mathbb E\log^+\left\vert  a_i  \right\vert <\infty$. 
Under the assumption that  $\pi_n(\mathbf A)$ has $d$ distinct Lyapunov exponents, $\forall$ $i$ $\exists \; X_0\in\mathbb C^d$ such that
\begin{eqnarray}
\lambda\left( \mathbf M ,  \left[ X_0^T \;\;1\right]^T   \right)=\max\left\{ \lambda_{\mathbf A,i} , \lambda_{a,1} \right\} \label{eq:Lx} .
\end{eqnarray}  
 \end{theorem}
 {
\begin{IEEEproof}
See Appendix \ref{apthrm1}.
 \end{IEEEproof}
 }

\subsubsection{ {Applications and/or Implications of Theorem \ref{lem:TTP}}}

We will now show that Theorem \ref{lem:TTP} can be used to calculate the Lyapunov  exponents of \eqref{eq:affine}. To do this, the affine structure of \eqref{eq:affine} will be captured by converting it into a linear (non-affine) $(d+1)\times (d+1)$ system of the following form:
\begin{equation}
 \left[\begin{array}{cc}
  X_n \\
 1
\end{array}\right] =  \left[\begin{array}{cc}
\mathbf A_n & R_n\\
\bf{0}^{T} & 1
\end{array}\right] \cdots  \left[\begin{array}{cc}
\mathbf A_1 & R_1\\
\bf{0}^{T} & 1
\end{array}\right] \left[\begin{array}{cc}
  X_{0} \\
 1
\end{array}\right]. \label{eq:nonaffine}
\end{equation}
One may naively assume that the Lyapunov exponents of \eqref{eq:affine} are trivially identical to those of \eqref{eq:nonaffine}. However, for this to be true, we must have  $\lim_{n\to\infty}\frac{1}{n}\log\left\Vert  X_n  \right\Vert \geq 0,$ because, clearly,
\begin{equation}
\lim_{n\to\infty}\frac{1}{n}\log\left\Vert \left[   X_n\;\;1  \right]  ^T \right\Vert \geq 0.\label{eq:SPNA}
\end{equation}
We now provide the following   lemma, which tells us that the Lyapunov exponents of \eqref{eq:affine} are indeed strictly non-negative, and that, consequently, the Lyapunov exponents of \eqref{eq:affine} and \eqref{eq:nonaffine} are identical.
\begin{lemma}\label{lem:AlwaysPos}
For $X_n$ given by \eqref{eq:affine}, we have
$$\lim_{n\to\infty}\frac{1}{n}\log||X_n|| \geq 0.$$
Consequently,  the Lyapunov exponents of the affine system, \eqref{eq:affine}, and those of the linear-to-affine converted system, \eqref{eq:nonaffine}, are identical.
\end{lemma}
\begin{IEEEproof}
See Appendix \ref{apAA}.
\end{IEEEproof}


 
From Fact \ref{lem:affine} (below), it can be seen that the Lyapunov exponents of   \eqref{eq:nonaffine} (and consequently \eqref{eq:affine}) must belong to  
$
\left\{ \max\left\{ \lambda_{\mathbf A,i} , 0 \right\}: i = 1,\dots ,d\right\}.$ 
 \begin{fact}  \label{lem:affine}
 \cite[Theorem 5]{key1988lyapunov}, consider the product of random matrices $\pi_n(\mathbf M)$, where
\begin{equation}\label{eq:Mmatrix}
\mathbf M_i =  \left[\begin{array}{cc}
\mathbf A_i & R_i\\
\bf{0}^{T} & a_i
\end{array}\right] \in\mathbb C^{(d+1)\times(d+1)},
\end{equation}
 $\mathbb E\log^+\left\Vert\mathbf A_i  \right\Vert <\infty$,   $\mathbb E\log^+\left\Vert  R_i  \right\Vert <\infty$, and $\mathbb E\log^+\left\vert  a_i  \right\vert <\infty$. 
Then the Lyapunov exponents of $\pi_n\left(\mathbf M\right)$ are given by $\{\lambda_{\mathbf{A},i}:i=1,\dots,n\}\cup\lambda_{a,1}$. Furthermore, the Lyapunov exponents of $\pi_n(\mathbf M)$ are independent of the statistics of $R_i$.
 \end{fact}

Notice that Fact \ref{lem:affine} tells us nothing about how initial states of the form $\left[  X_0^T \; \;1  \right]^T$ (c.f. \eqref{eq:nonaffine}), will affect the Lyapunov analysis. Consequently, we require a theorem that deals with such initial states. This provides the rationale behind Theorem~\ref{lem:TTP}. It is important to mention that Theorem~\ref{lem:TTP} also plays an important role within the proof of Theorem~\ref{lem2}.
%
%


 \section{  { Capacity and Power       Scaling\label{sec:PwrCap}} }

In this section, we will use the network's Lyapunov exponents to establish the scaling behavior of the end-to-end capacity  and $n$th node  transmit power. 
The following theorem  relates the  Lyapunov exponents to the network's end-to-end capacity.
\begin{theorem}\label{lem2}Let $\lambda_{\alpha\mathbf{H},i}$ be the Lyapunov exponent of $\mathcal I_n^{(\alpha)}$ (which is defined in \eqref{eq:SISO_Infor_Form}) and  $ 
c_n^{(\alpha)}  = \sum_{i=1}^{d}c_{n,i}^{(\alpha)}\;\mathrm{(nats / channel\; use)}
$ be the capacity of the $n$-hop AF network with amplification strategy $\alpha$ (see Definition \ref{def:capacity_nhop} and Fact \ref{fact:capacity_nhop}), where
$
c_{n,i}^{(\alpha)}
$
is the capacity of the $i$th eigenchannel at the $n$th node. Then the following statements hold.
\begin{description}
\item[\emph{A.}]  
$
\lim_{n\to\infty}\frac{1}{n}\log\mathcal E_i\left(\mathbf R_{\mathcal{I},n}^{(\alpha)}\mathbf R_{\mathcal{N},n}^{(\alpha)-1} \right) \overset{a.s.}{=} \min\{0, 2\lambda_{\alpha\mathbf{H},i}  \}  =: \lambda_{\gamma,i}^{(\alpha)}. \label{eq:SV_Lya}
$
Hence, the SNR of the $i$th eigenchannel obeys
\begin{equation}
\mathcal E_i\left(\mathbf R_{\mathcal{I},n}^{(\alpha)}\mathbf R_{\mathcal{N},n}^{(\alpha)-1} \right) = \Theta_{\mathbb P}\left( e^{n  \lambda_{\gamma,i}^{(\alpha)} }\right)\label{eq:EigEquiv}.
\end{equation}
\item[\emph{B.}]$
\lim_{n\to\infty}\frac{1}{n}\log c_{n,i}^{(\alpha)} \overset{a.s.}{=} \min\{0, 2\lambda_{\alpha\mathbf{H},i}  \}  =: \lambda_{\gamma,i}^{(\alpha)} . \label{eq:Cap_Lya}
$
Thus, the capacity of the $i$th eigenchannel obeys
\begin{equation}
c_{n,i}^{(\alpha)} = \Theta_{\mathbb P} \left(   e^{n\lambda_{\gamma,i}^{(\alpha)}}  \right).\label{eq:asmC}
\end{equation}
\end{description}
\end{theorem}
\begin{IEEEproof}
See Appendix \ref{apB}.
\end{IEEEproof}

The next lemma will evaluate the Lyapunov exponent $\lambda_{\alpha\mathbf{H} , i}$, and establish how it  relates to the Lyapunov exponents of $X_n^{(\alpha)}$ and the average transmit power at the $n$th node. This lemma will in turn allow us  to establish a trade off between capacity decay and power growth across the network. It will also have implications on gain design.
\begin{lemma}\label{lem:AVGPWR}With $\mathcal I_n^{(\alpha)}$ given by \eqref{eq:SISO_Infor_Form}, $X_n^{(\alpha)}$ given by \eqref{eq:x}, and the average transmit power at the $n$th node given by $p_n$, the following statements hold.
\begin{description}
\item[\emph{A.}]   The $i$th Lyapunov exponent of $\mathcal I_n^{(\alpha)}$ is given by
\begin{equation}
\lambda_{\alpha\mathbf{H},i} = \frac{1}{2}\left( L\left(   {\alpha}^2\mu \right) + \psi(d-i+1)\right), \; i=1,\!\cdots\!,d;\label{eq:LI}
\end{equation}
where $\psi\left( \cdot \right)$ is the digamma function \cite[eq. (6.3.1)]{abramowitz1964handbook} and 
\begin{equation}
L(\alpha^2\mu) :=  \lim_{n\to\infty}\frac{1}{n}\sum_{i=1}^n \log( \alpha_i^2\mu_i).\label{eq:L}
\end{equation}
The $i$th Lyapunov exponent of $X_n^{(\alpha)}$  is given by
\begin{equation}
\lambda_{\mathbf{ Q},i}^{(\alpha)}  :=  \lambda\left(  \mathbf Q, \left[ X_0^T\;\sqrt{n_0} \right]  \right) = \max\left\{  0, \lambda_{\alpha\mathbf{H},i}   \right\}.
\end{equation} Hence, the  information power and total transmit power obey
\begin{eqnarray}
\left\Vert\mathcal I_n^{(\alpha)}\right\Vert^2& =&   \Theta_{\mathbb P}\left(   e^{2n\lambda_{\alpha\mathbf{H},1} }\right),\label{eq:Itraj}\\
\left\Vert X_n ^{(\alpha)}\right\Vert^2 &=& \Theta_{\mathbb P}\left(  e^{2n\lambda_{\mathbf Q,1}^{(\alpha)}}\right)\label{eq:xtraj}.\label{eq:Xtraj}
\end{eqnarray}
\item[\emph{B.}]For fixed-gain, we have
\begin{equation}
 \lim_{n\to\infty}\frac{1}{n}\log \frac{p_n}{p_0}  \geq \max\left\{ 2\lambda_{f\mathbf{H},1}  +  \log d - \psi( d ) ,\; 0 \right\};\label{eq:FGPB}
\end{equation}
for variable-gain, we have
\begin{multline}
 \lim_{n\to\infty}\frac{1}{n}\log \frac{p_n}{p_0} \geq \max\left\{ 2\lambda_{v\mathbf{H},1} +\psi( d^2) \right. \\ \left. -\psi(d)- \log d ,\; 0 \right\};\label{eq:VGPB}
\end{multline}
where  equality is maintained only when $n_0=0$.
\end{description}
\end{lemma}
\begin{IEEEproof}
See Appendix \ref{ap:AVG_Grow}. 
\end{IEEEproof}

\subsection{ {A Brief Discussion of  Theorem \ref{lem2} and Lemma \ref{lem:AVGPWR} }}

%

From the first statement of Lemma \ref{lem:AVGPWR},  
by ensuring 
$
\lambda_{\alpha\mathbf{H},d}  <   \cdots   < \lambda_{\alpha\mathbf{H},1} =0,
$
we can avoid exponential growth in the instantaneous transmit power.
However, in this setup Theorem \ref{lem2} tells us that all but the first eigenchannel will display an exponentially decaying capacity. 
Conversely, by ensuring 
$
\lambda_{\alpha\mathbf{H},1}  >\cdots >\lambda_{\alpha\mathbf{H},q}  \geq 0 > \lambda_{\alpha\mathbf{H}, q+1} > \cdots > \lambda_{\alpha\mathbf{H}, d} ,
$
we can stop the end-to-end capacity of the upper $q$ eigenchannels from \emph{almost surely} decaying exponentially. However, in this scenario, we must allow for exponential  growth in the instantaneous power across the network.
Thus, there is a clear tradeoff to be had between multiplexing multiple data streams across the network, and growth in the instantaneous transmit power at the $n$th node.

Focusing on the second statement of Lemma \ref{lem:AVGPWR}, it can be seen that the terms  $\log d-\psi(d)$ and $\psi( d^2)-\psi(d)-\log d$ in \eqref{eq:FGPB} and \eqref{eq:VGPB}, respectively, are strictly non-negative.
 Thus,  this statement  tells us that, asymptotically, the \emph{average} transmit power must grow at a greater exponential rate than the instantaneous power. 
Crucially, we find that exponential growth in $p_n$ can be allowed for whilst avoiding (with high probability) exponential instantaneous  power growth at the relays. Said in a different way, as the network scales in size, the density function of the transmit power at the $n$th node becomes increasingly heavy tailed. Whilst most of the distribution's mass will be concentrated at the point governed by the Lyapunov exponent (cf. \eqref{eq:Xtraj}), the distribution's heavy tail will push the average up exponentially. Combining this observation with Theorem \ref{lem2}, it can be seen that ensuring  the first eigenchannel displays a non exponentially decaying capacity implies that the average transmit power will grow exponentially.

It can also be seen that,  because  $\log d-\psi(d)\geq\psi( d^2)-\psi(d) - \log d$, the lower bound on the exponential growth rate of the average transmit power for variable-gain is strictly less than that for fixed-gain, which suggests that the variable-gain network can sustain an approximately constant instantaneous power trend with a reduced growth in the average transmit power. Furthermore, as the number of antennas grows large, both bounds in Lemma~\ref{lem:AVGPWR} converge towards the Lyapunpov exponents. Thus, ergodic behavior is induced as $d$ grows large. 

In summary, Theorem \ref{lem2} and the first statement of Lemma \ref{lem:AVGPWR} expose a fundamental trade off between capacity decay and instantaneous transmit power growth across the network. The second statement of Lemma \ref{lem:AVGPWR}  has  important implications  on gain design for scaled networks. In particular, it implies that the average transmit power at each node should grow exponentially with the network if an approximately constant instantaneous power trend is to be maintained. These implications contrast with the system model proposed in \cite{5074422,6169203}, where the capacity was assessed under strictly linear scaling of $p_n$. For the finite antenna system, we see that if linear scaling of $p_n$ occurs, $\lim_{n\to\infty}\log(p_n/p_0)/n=0$  and  (from Lemma \ref{lem:AVGPWR}) $\lambda_{\alpha\mathbf{H},1} <0$. 
As has been seen in the Theorem \ref{lem2}, $\lambda_{\alpha\mathbf{H},1} < 0$ will have serious implications on the network's end-to-end capacity.  As an extra note, it can be seen that our result implicitly applies to a network whose length grows with the number of hops in the network (the extended regime); i.e., the distance between each node is fixed. For future work, it may be interesting to consider capacity and power scaling properties for networks when the end-to-end length of the network is fixed, and the distance between each of the nodes decreases with the number of hops (the dense regime), see \cite{ozgur2007hierarchical}. However, this is beyond the scope of this manuscript. 

 Finally, the authors would like to point out that it is unclear whether our RDS and corresponding Lyapunov analysis can be applied to study other forwarding schemes (e.g., DF).  For the AF scenario, the analysis relies on the ability to make a correspondence between the network and products of random matrices. Moving to other (non AF) forwarding scenarios, the relay network will be described by a composition of random nonlinear mappings. In general, when determining the Lyapunov exponents of a system, the multiplicative ergodic theorem \cite{arnold1998random} (MET) is referred to. This theorem is a linear result, and when one refers to the MET for a nonlinear system they are implicitly referring to the application of this theorem to the linearized version of the nonlinear system. Thus, we have two open questions about applying our approach to other forwarding schemes:
 \begin{enumerate}
\item Does the linearization of a system describing forwarding such as DF make sense from a practical view point?
\item If the answer to 1) is yes, can analogous capacity and power  results  be obtained for such schemes? 
\end{enumerate}

\section{ {Applications of Theorem \ref{lem2} and Lemma \ref{lem:AVGPWR} \label{sec:ECapDiv}}}

In this section, we will study some  applications of  Theorem~\ref{lem2} and Lemma \ref{lem:AVGPWR}. In particular, we will study the rates at which the eigenchannel capacities diverge away from each other, and how this relates to:  
\begin{itemize}
\item  the amplification strategy and   number of antennas at each node, 
  \item the  growth in the instantaneous transmit power.
  \end{itemize}
    To discuss the above mentioned points, we will require the following preliminary definitions and lemmas.

 {\subsection{Preliminary Definitions and Lemmas}}

\begin{definition} \label{def:v}The $(i,j)$th normalized channel capacity, $i\leq j$, is defined to be
\begin{equation}
\nu_{i,j,n}^{(\alpha)} := \frac{c_{i,n}^{(\alpha)} }{c_{j,n}^{(\alpha)}}\label{eq:Cnum}.
\end{equation}
\end{definition} 

Clearly, if $\nu_{1,j,n}^{(\alpha)}\approx1$, the channel will be well suited for multiplexing $j$ data streams, \cite{tse2005fundamentals,goldsmith2005wireless}, provided $c_{1,n}^{(\alpha)}$ is sufficiently large; otherwise, it will not. 

 \begin{definition}\label{def:LyapDiff}For both fixed-gain and  variable-gain, the $(i,j)$th \emph{Lyapunov difference},   $i\leq j$, is defined to be
\begin{equation}
\phi_{i,j}^{(\alpha)} :=  \lambda_{\gamma  ,i}^{(\alpha)}  - \lambda_{\gamma  ,j}^{(\alpha)} .\label{eq:LyaSpread}
\end{equation}
\end{definition}

The following two lemmas are used to bound $\phi_{i,j}^{(\alpha)}$, and will be employed in the ensuing  analysis.
 \begin{lemma}\label{lem9}
 The $(i,j)$th Lyapunov difference is  bounded as follows:
 \begin{equation}
0\leq \phi_{i,j}^{(\alpha)} \leq 2\left(\lambda_{\alpha\mathbf{H},i}  - \lambda_{\alpha\mathbf{H},j} \right) =:\bar\phi_{i,j} , \label{eq:spreadUB}
 \end{equation}
 where lower equality is maintained if and only if $\lambda_{\alpha\mathbf{H},i}>\lambda_{\alpha\mathbf{H},j}\geq 0$, upper equality is maintained if and only if $\lambda_{\alpha\mathbf{H},j}<\lambda_{\alpha\mathbf{H},i}\leq 0$, and $\phi_{i,j}^{(\alpha)} = -2\lambda_{\alpha\mathbf{H},j}$ otherwise. Furthermore, the upper bound is indepedent of whether fixed-gain or variable-gain is being implemented.
 \end{lemma}
 \begin{IEEEproof}
 See Appendix \ref{spreadUB}.
 \end{IEEEproof}
 
 Finally, we will also exploit the following lemma later in this section.
\begin{lemma}\label{lem:harm1}
For $i<j$, we have
 \begin{eqnarray}
 \bar\phi_{i,j}& = &  \sum_{k=d-j+1}^{d-i} \frac{1}{k}  =  \left( \mathcal H_{d-i}   -  \mathcal H_{d-j}   \right)  \label{eq:Ispread},
 \end{eqnarray}
 where $\mathcal H_i$ is the $i$th harmonic series defined to be
\begin{equation}
\mathcal H_i = \sum_{j=1}^{i}\frac{1}{j}.
\end{equation}
Furthermore, by considering the first and last summands in \eqref{eq:Ispread}, we can trivially construct the following bound:
\begin{equation}
\frac{j-i+1}{d-i} \leq \bar\phi_{i,j}  \leq \frac{j-i+1}{d-j+1}.
\end{equation}
\end{lemma}
\begin{IEEEproof}
This follows immediately from \eqref{eq:LI}, Lemma \ref{lem9}, and applying the telescope property of the digamma function \cite[eq. (6.3.5)]{abramowitz1964handbook}:
\begin{equation}
\psi(x+1) = \psi(x) +\frac{1}{x}.\label{eq:telprop}
\end{equation}
\end{IEEEproof}


 {\subsection{Growth of $\nu_{i,j,n}^{(\alpha)}$ and $\Vert X_n\Vert^2$}}

We will now apply  Theorem \ref{lem2} and Lemma \ref{lem:AVGPWR} to study $\nu_{i,j,n}^{(\alpha)}$ (Definition \ref{def:v}) and $\Vert X_n^{(\alpha)} \Vert^2$.  Considering Theorem~\ref{lem2} first, from Definitions~\ref{def:v} and \ref{def:LyapDiff} we have
\begin{equation}
\lim_{n\to\infty}\frac{1}{n}\log \nu_{i,j,n}^{(\alpha)}=  \phi_{i,j}^{(\alpha)} \Longleftrightarrow\nu_{i,j,n}^{(\alpha)}= \Theta_{\mathbb P}\left( e^{ n \phi_{i,j}^{(\alpha)}  }\right).\label{eq:CnumLya}
\end{equation}
We will now use \eqref{eq:CnumLya} (in conjunction with Lemma \ref{lem:AVGPWR}) to study the following  four problems:
\begin{enumerate}
\item The dependence of the growth in $\nu_{i,j,n}^{(\alpha)}$ on the amplification strategy.
\item The dependence  of the growth in $\nu_{i,j,n}^{(\alpha)}$ on the number of antennas at each node
\item The behavior of the network when either $\phi_{1,i}^{(\alpha)} = 0$ or $\lambda_{\alpha\mathbf{H},1}= 0$; i.e., when either  $\nu_{i,j,n}^{(\alpha)}$  or $\Vert X_n^{(\alpha)}\Vert^2$ display no exponential growth, respectively.
\item The growth in $\nu_{i,i+1,n}^{(\alpha)}$ (i.e., rate at which adjacent eigenchannel capacities diverge away from each other), and the cost (in terms of instantaneous transmit power) associated with each extra multiplexed data stream.
\end{enumerate}

\subsubsection{ {Growth of $\nu_{i,j,n}$ and the Forwarding Strategy }\label{sec:fwrdstrat}}

  Let us first establish how the amplification strategy affects the growth of $\nu_{i,j,n}^{(\alpha)}$. As an immediate consequence of Lemma~\ref{lem9}, it can be seen that  when $\lambda_{\alpha\mathbf{H},j}<\lambda_{\alpha\mathbf{H},i}\leq 0$ the exponential growth of $\nu_{i,j,n}^{(\alpha)}$ will be independent of the amplification strategy that has been implemented. The same holds true when $0\leq\lambda_{\alpha\mathbf{H},j}<\lambda_{\alpha\mathbf{H},i} $, since we will have $\phi_{i,j,n}^{(\alpha)} = 0$. For $ \lambda_{\alpha\mathbf{H},j}< 0 < \lambda_{\alpha\mathbf{H},i} $, we will have $\phi_{i,j,n}^{(\alpha)} = -2\lambda_{\alpha\mathbf{H},j}$. Consequently, in this scenario $\nu_{i,j,n}^{(\alpha)}$ is given by
\begin{align}
\nu_{i,j,n}^{(f)}  &= \Theta_\mathbb{P} \left(  e^{ -2\lambda_{f\mathbf{H},j}}  \right)\nonumber\\ &= \Omega_\mathbb{P} \left(  e^{- \lim_{n\to\infty} \frac{1}{n}\log\frac{p_n}{p_0}+\log d - \psi(d-j+1)}  \right),\label{eq:vijnFG}
\end{align}
for fixed-gain,
   and 
\begin{align}
\nu_{i,j,n}^{(v)} &= \Theta_\mathbb{P} \left(  e^{ -2\lambda_{v\mathbf{H},j}}  \right)\nonumber \\&= \Omega_\mathbb{P} \left(  e^{ - \lim_{n\to\infty} \frac{1}{n}\log\frac{p_n}{p_0} -  \log(d)+ \psi\left( d^2 \right) - \psi\left( d- j +1 \right)  }  \right),\label{eq:vijnVG}
\end{align}
 for  variable-gain, where the second equalities of \eqref{eq:vijnFG} and \eqref{eq:vijnVG} follow  from Lemma \ref{lem:UB} (see Appendix \ref{ap2}).  
Notice that, because $  \log d^2 >   \psi \left(d^2\right)$, 
\begin{multline}
e^{- \lim_{n\to\infty} \frac{1}{n}\log\frac{p_n}{p_0}+\log d - \psi(d-j+1)}\\ \geq e^{ - \lim_{n\to\infty} \frac{1}{n}\log\frac{p_n}{p_0} -  \log(d)+ \psi\left( d^2 \right) - \psi\left( d- j +1 \right)  }.\label{eq:FGgtVG}
\end{multline}
 \subsubsection{ {Growth of $\nu_{i,j,n}$  and the  Number of Antennas}}
We will now establish how the number of antennas at each node will affect the growth rate of $\nu_{i,j,n}^{(\alpha)}$. In particular, we will determine how $n\phi_{i,j}^{(\alpha)}$ (the term in the exponent of \eqref{eq:CnumLya}) scales with $n$, and how the number of antennas relates to this. More specifically, in what follows (Lemma \ref{lem:1stOrder} and Corollary \ref{cor:9}) we will determine conditions that give the following:
  \begin{eqnarray}
 \lim_{n\to\infty}\left[ n\phi_{i,j}^{(\alpha)}\right] =  0\;\;&\Leftrightarrow& \phi_{i,j}^{(\alpha)} = o\left(  1/n \right) \label{item:limnphi0},\\
 \!\!\!\!\!\!\!\!\!\!0<\lim_{n\to\infty}\left[ n\phi_{i,j}^{(\alpha)}\right] \leq k < \infty&\Leftrightarrow& \phi_{i,j}^{(\alpha)} = \Theta\left(  1/n \right) ,\label{item:limnphiK}\\
 \lim_{n\to\infty}\left[ n\phi_{i,j}^{(\alpha)}\right] =  \infty\;\!&\Leftrightarrow& 1 / \phi_{i,j}^{(\alpha)} = o\left(   n  \right). \label{item:limnphiinf}
 \end{eqnarray}
Of course, if $\phi_{i,j}^{(\alpha)} = 0$ (i.e., $\lambda_{\alpha\mathbf{H},i} > \lambda_{\alpha\mathbf{H},j}\geq 0$)    \eqref{item:limnphi0} is obtained trivially. We are therefore only interested in studying the behavior of $n\phi_{i,j}^{(\alpha)} $ when either   $0\geq \lambda_{\alpha\mathbf{H},i}  >\lambda_{\alpha\mathbf{H},j}$ or  $\lambda_{\alpha\mathbf{H},i}  > 0 >\lambda_{\alpha\mathbf{H},j}$. We treat $0\geq \lambda_{\alpha\mathbf{H},i}  >\lambda_{\alpha\mathbf{H},j}$  in Lemma~\ref{lem:1stOrder} and consider $\lambda_{\alpha\mathbf{H},i}  > 0 >\lambda_{\alpha\mathbf{H},j}$ in its corollary.
\begin{lemma}\label{lem:1stOrder}
When $0\geq \lambda_{\alpha\mathbf{H},i}  >\lambda_{\alpha\mathbf{H},j}$, for both fixed-gain and  variable-gain, to leading order about $d =\infty$ (i.e., $i<j$ fixed and $d\to\infty$)    $\phi_{i,j}^{(\alpha)}$ is given by
\begin{equation}
\phi_{i,j}^{(\alpha)} = \bar\phi_{i,j}= \frac{ (j-i)}{d} + O\left( \frac{1}{d}\right)^2  .  \label{eq:1orderphi}
 \end{equation}
 Consequently,  
\begin{enumerate}
\item the conditions that give  \eqref{item:limnphi0} are $ 1/d= o(1/n )  $,
\item the conditions that give  \eqref{item:limnphiK} are $ 1/d= \Theta(1/n )  $
\item the conditions that give  \eqref{item:limnphiinf}  are $ d= o(n )  $.
 \end{enumerate}
 \end{lemma}
  \begin{IEEEproof}
Eq. \eqref{eq:1orderphi}  is obtained by performing a Taylor expansion of $\bar\phi_{i,j}$ about the point $d$ and letting $d\to\infty,$ with $i\leq j$ fixed. The following statements  then follow immediately.
 \end{IEEEproof}
\begin{corollary}\label{cor:9}
It is only possible to maintain $\lambda_{\alpha\mathbf{H},i}  > 0 >\lambda_{\alpha\mathbf{H},j}$ when $d = O(1)$. From Lemma~\ref{lem:1stOrder}, when this occurs $\lim_{n\to\infty} n \phi_{i,j}^{(\alpha)} = \infty$.
\end{corollary}

We will now discuss Lemma \ref{lem:1stOrder} and its corollary. These are seen to complement \cite[Theorem. 4]{6169203}, in which it was shown that $\lim_{n\to\infty}\left[\lim_{d_D\to\infty}\left[c_n/d_D\right]/n\right]$ (where $d_D$ is the number of destination antennas) will be strictly positive if and only if $d/d_D = \Theta (n^{1+\epsilon})$ for all $\epsilon \geq 0$ (note, the inequality for $\epsilon$ is not strict).  In   our work, if $d/j= \Theta (n^{1+\epsilon})$, for fixed $j$, $n\phi_{i,j}^{(\alpha)}$ will be bounded away from infinity $\forall\;i<j$ and consequently, from \eqref{eq:CnumLya}, $\nu_{i,j,n}^{(\alpha)}$ will \emph{almost surely} display no exponential growth as $n$ grows without bound\footnote{For the work in \cite{6169203} and our work, $d_D$ and $j$ can be thought of as the maximum number of data streams that can be multiplexed over the channel, respectively. This draws the connection between that work, where the scaling of the ratio $d/d_D$ is assessed, and our work, where the scaling of $d/j$ is assessed.}. Clearly, avoiding exponential growth of $\nu_{1,j,n}^{(\alpha)}$ is required if we are to multiplex over the $j$ upper eigenchannels.  Crucially,  these results provide us with an alternative perspective to \cite{6169203} on how the number of antennas (more precisely, the scaling of this number) at each node affects the end-to-end capacity of the network.

\subsubsection{ {Network behavior when $\phi_{1,i}^{(\alpha)} = 0$ or $\lambda_{\alpha\mathbf{H},1}= 0$}}
 
Suppose   we wish to ensure that the $(1,i)$th normalized channel capacity  displays no exponential growth; i.e.,  (from \eqref{eq:CnumLya})  $\phi_{1,i}^{(\alpha)} = 0$. Furthermore, suppose this is  achieved by ensuring that
\begin{equation}
 \lambda_{\alpha\mathbf{H},1}>\lambda_{\alpha\mathbf{H},i}= 0.
\end{equation}
Then \eqref{eq:xtraj} and Lemma \ref{lem:harm1} give us
\begin{equation}
\left\Vert X_n^{(\alpha)}  \right\Vert^2 = \Theta_{\mathbb P} \left( e^{n(\mathcal H_{d-1} - \mathcal H_{d-i})}\right),\label{eq:X_nH}
 \end{equation} 
 where the argument of $ \Theta_{\mathbb P} \left(\cdot\right)$ in  \eqref{eq:X_nH} is bound in the following way:
 \begin{equation}
  e^{\frac{ni}{d-1}} \leq   e^{n(\mathcal H_{d-1} - \mathcal H_{d-i})}\leq   e^{ \frac{ni}{d-i+1} }\label{eq:TPbound}.
 \end{equation} 
Thus, ensuring     $\phi_{1,i}^{(\alpha)} = 0$ implies that the transmit power must grow according to \eqref{eq:X_nH}. This  growth rate is strictly positive and bound according to \eqref{eq:TPbound}. We can see that by increasing the number of antennas, $d$, for a fixed $i$, the rate at which the transmit power grows can be reduced. Conversely, by fixing $d$ and increasing $i$ (i.e., multiplexing more data streams), the rate at which the transmit power must grow will increase. 

Suppose instead   we wish to ensure that the transmit power displays no exponential growth by setting $\lambda_{\alpha\mathbf{H},1} = 0$. From   \eqref{eq:CnumLya} and Lemmas \ref{lem9} and \ref{lem:harm1}, this gives
\begin{equation}
\nu_{1,i,n}^{(\alpha)}= \Theta_{\mathbb P} \left(  e^{n(\mathcal H_{d-1} - \mathcal H_{d-i})}   \right).   \label{eq:CN_nH}
\end{equation}
Thus, all of the growth properties  that applied to $\Vert X_n^{(\alpha)}\Vert^2$ when $\lambda_{\alpha\mathbf{H},i} = 0$ apply to $\nu_{1,i,n}^{(\alpha)}$ when $\lambda_{\alpha\mathbf{H},1} = 0$.
\begin{remark}\label{rem:duality}
Interestingly, from \eqref{eq:X_nH} and \eqref{eq:CN_nH}, it can be seen that there is a duality between the exponential growth rate of $\left\Vert X_n^{(\alpha)}\right\Vert^2$ and $\nu_{1,i,n}^{(\alpha)}$ when either $\lambda_{\alpha\mathbf{H},i} = 0$ or $\lambda_{\alpha\mathbf{H},1} = 0$, respectively. This duality property will be exploited below.
\end{remark}

\subsubsection{ {Adjacent Eigenchannel Capacity Divergence and Individual Data Stream Cost}}

For the final problem, let us consider the rate at which adjacent eigenchannel capacities  diverge away from each other. Of course, we have already seen (Lemma \ref{lem9} and \eqref{eq:CnumLya}) that if $\lambda_{\alpha\mathbf{H}, i}>\lambda_{\alpha\mathbf{H},i+1}\geq 0$ then $c_{i,n}$ and $c_{i+1,n}$ will not diverge away from each other. Thus, in what follows we consider the cases  $0 \geq \lambda_{\alpha\mathbf{H}, i} > \lambda_{\alpha\mathbf{H},i+1} $ and $\lambda_{\alpha\mathbf{H}, i}> 0 > \lambda_{\alpha\mathbf{H},i+1} $.

When $0\geq\lambda_{\alpha\mathbf{H}, i} > \lambda_{\alpha\mathbf{H},i+1} $, by employing Lemma  \ref{lem:harm1} we find that 
\begin{equation}
\nu_{i,i+1,n}^{(\alpha)}
=  \Theta_{\mathbb P} \left( e^{\frac{ n}{d-i}} \right) \label{eq:eigdiv2}.
\end{equation}
Thus, the  $i$th and $(i+1)$th channel capacities diverge away from each other at an exponential rate ${1}/{(d-i)}$.
When $\lambda_{\alpha\mathbf{H}, i}\geq 0 > \lambda_{\alpha\mathbf{H},i+1} $ we find that 
\begin{equation}
\nu_{i,i+1,n}^{(\alpha)}
=\Theta_{\mathbb P} \left( e^{-2n \lambda_{\alpha\mathbf{H},i+1}}\right) = O_{\mathbb P}\left(   e^{\frac{ n}{d-i}}   \right)\label{eq:eigdiv}
\end{equation}
and the capacities diverge away from each other at an exponential rate $-2 \lambda_{\alpha\mathbf{H},i+1}$, which  is upper bounded by the exponential rate of \eqref{eq:eigdiv2}. 

\begin{remark}\label{rem:cost}
By considering the discussion of duality in Remark \ref{rem:duality}, we can assign a cost (in terms of extra instantaneous power requirements) to each extra data stream that we attempt to multiplex. In particular,    from \eqref{eq:eigdiv2} and because of the duality property, if we are multiplexing $i$ data streams, then, to multiplex $1$ more stream (whilst ensuring $\lambda_{\alpha\mathbf H,i+1}=0$), we must increase the $n$th relay's instantaneous transmit power by (approximately) a factor of $\exp(n/(d-i))$. Furthermore, we find that the cost of each extra eigenchannel increases with $i$.
 \end{remark}

  {\section{Numerical Illustration\label{sec:Illus}}}
  
In this section, we will illustrate the theory that has been presented in the previous sections. It is important to mention firstly that the following Monte Carlo simulations were generated using the variable precision arithmetic (vpa) function within the Matlab symbolic toolbox, which allowed us to increase the accuracy of our calculations to (approximately) $100$ decimal places. This is required because of the nature of our results: we are verifying as clearly as possible that the eigenchannel capacities follow exponential trends governed by their corresponding Lyapunov exponents. For large networks, this results in computational rounding within the simulations if vpa is not utilized. An immediate consequence of employing such high precision is that simulations are \emph{very} computationally intensive. Crucially, this restricts us to demonstrating network trends when the number of antennas at each node are small (i.e., $3$ or $4$). 

\begin{figure}
\centering
\includegraphics[scale=0.62]{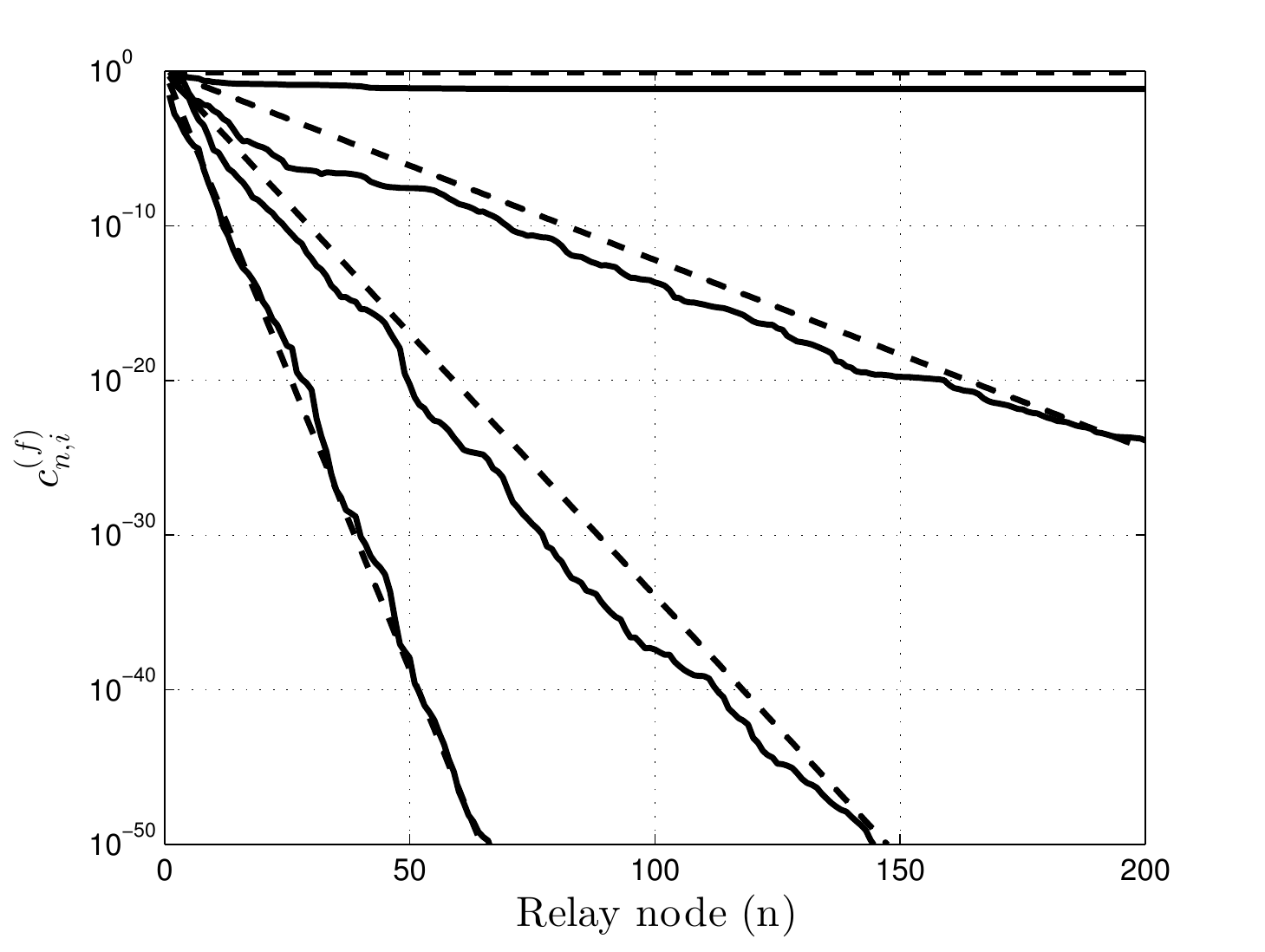}\caption{\label{fig:Cap_FG}Figure demonstrating that the capacity of each eigenchannel is given by $c_{n,i}^{(f)}=\Theta_{\mathbb P}\left( e^{n\lambda_\gamma^{(f)}} \right)$ for a $4\times 4$ fixed-gain MIMO system, see Theorem   \ref{lem2}. Dashed lines represent  $\exp\left(n\lambda_{\gamma,i}^{(f)}\right)$; solid lines represent instantaneous realizations of $c_{n,i}^{(f)},$ where, starting from the top, $i=1,\cdots,4.$  For all $i=1,\dots,n$, we set $p_i=n_0=\mu_i = 1.$}
\end{figure}

\begin{figure}
\centering
\includegraphics[scale=0.62]{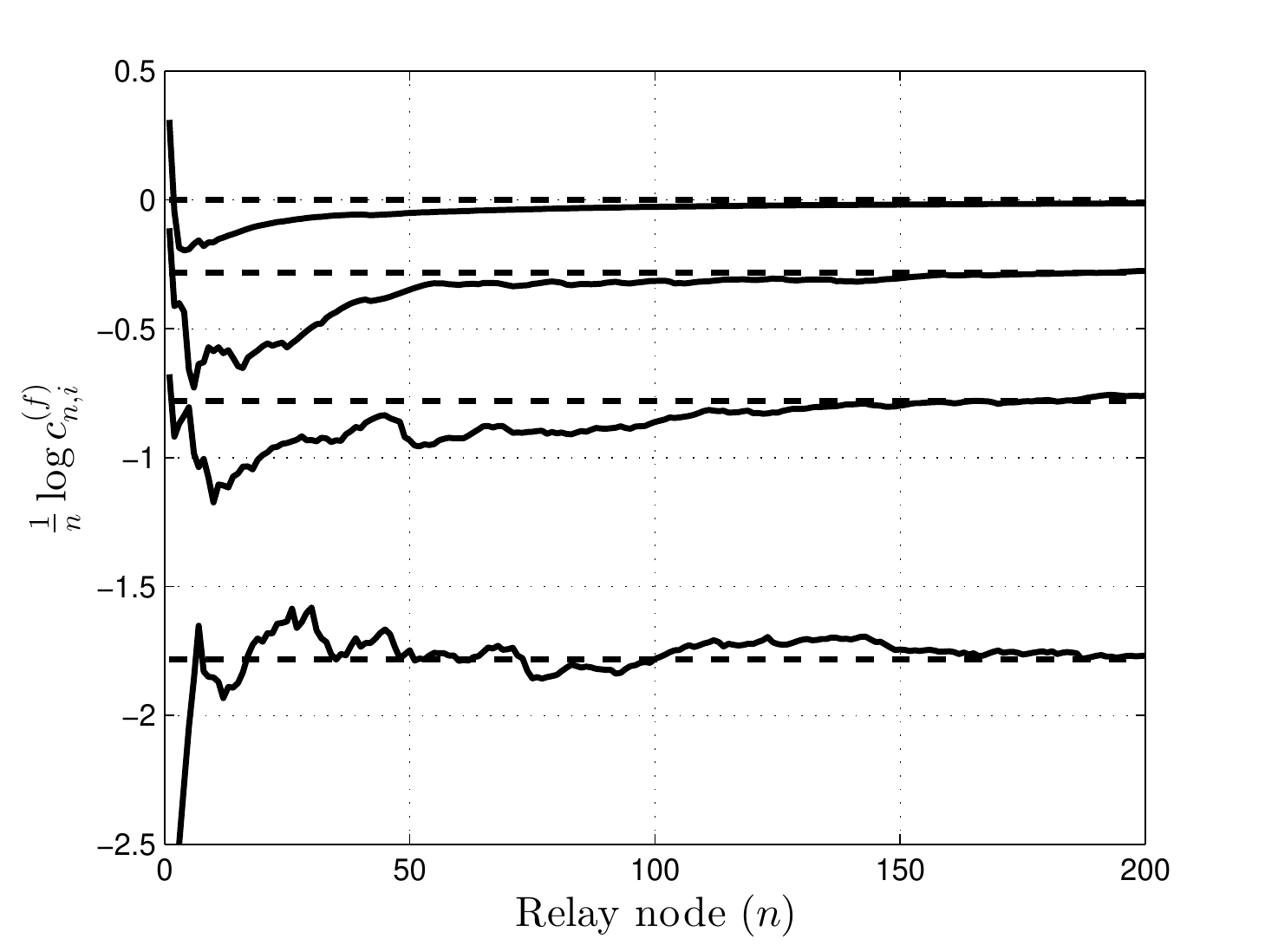}\caption{\label{fig:Cap_Lya} Figure demonstrating the second statement of Theorem   \ref{lem2} for a $4\times 4$ fixed-gain MIMO system. Dashed lines represent the Lyapunov exponents $\lambda_{\gamma,i}^{(f)}$, \eqref{eq:SV_Lya}; solid lines represent instantaneous realizations of $\frac{1}{n}\log  c_{n,i}^{(f)},$ where, starting from the top, $i=1,\cdots,4.$  For all $i=1,\dots,n$, we set $p_i=n_0=\mu_i = 1.$}
\end{figure}

\begin{figure}
\centering
\includegraphics[scale=0.62]{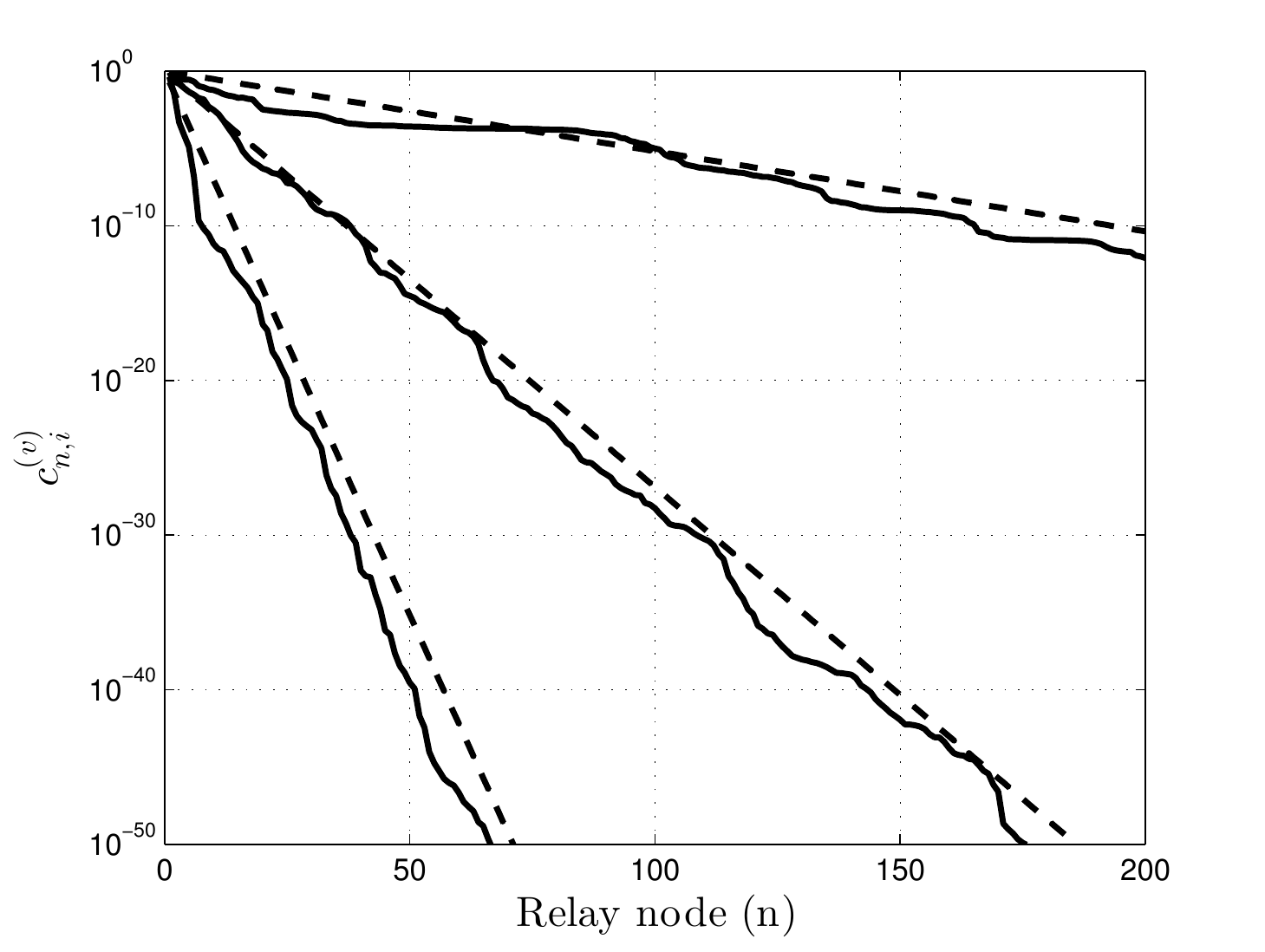}\caption{\label{fig:Cap_VG}Figure demonstrating that the capacity of each eigenchannel is given by $c_{n,i}^{(v)}=\Theta_{\mathbb P}\left( e^{n\lambda_\gamma^{(v)}} \right)$ for a $3\times 3$ variable-gain MIMO system, see Theorem   \ref{lem2}. Dashed lines represent  $\exp\left(n\lambda_{\gamma,i}^{(v)}\right)$; solid lines represent instantaneous realizations of $c_{n,i}^{(v)},$ where, starting from the top, $i=1,\cdots,3.$  For all $i=1,\dots,n$, we set $p_i=n_0=\mu_i = 1.$}
\end{figure}

\begin{figure}
\centering
\includegraphics[scale=0.62]{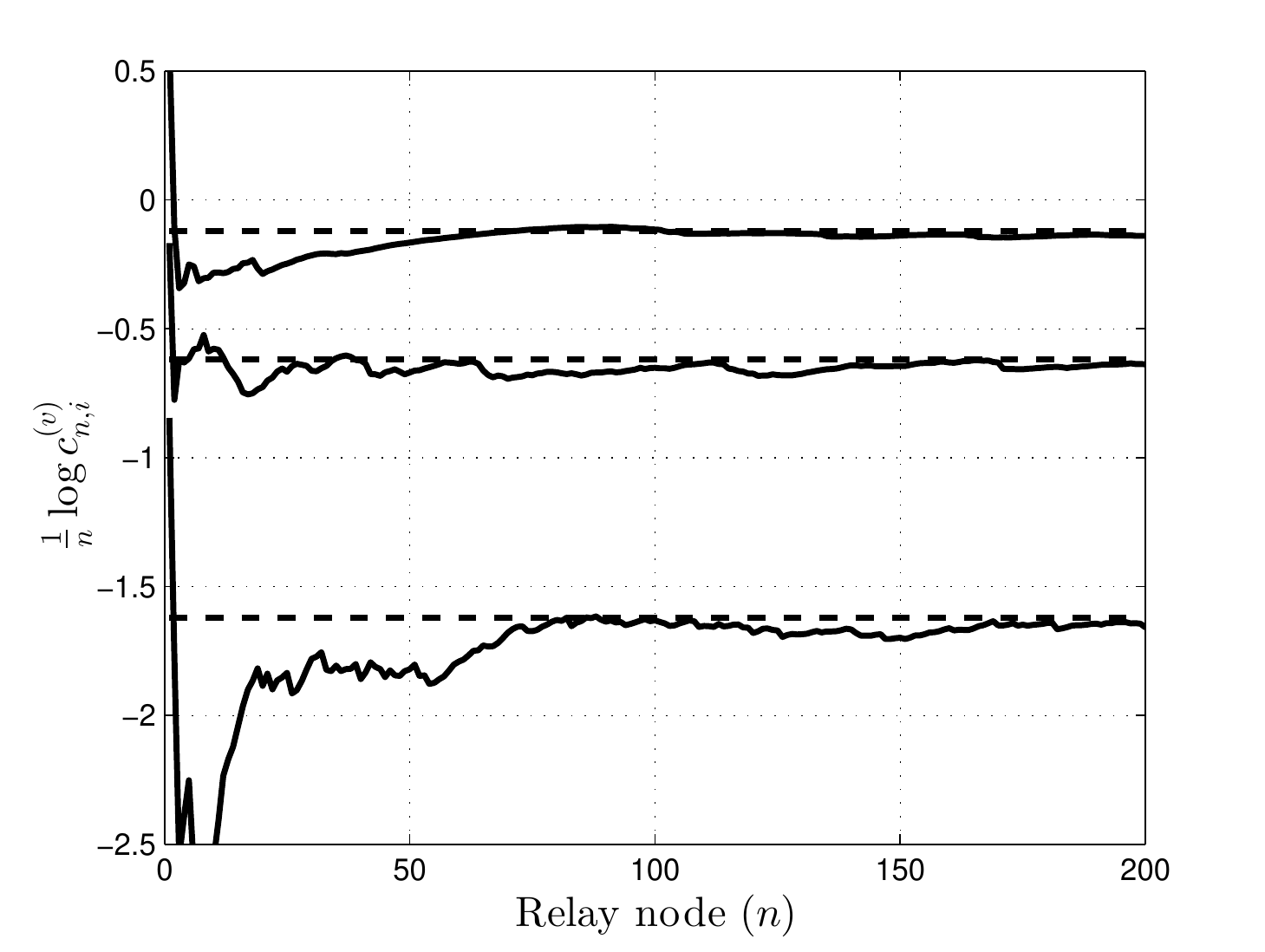}\caption{Figure demonstrating the second statement of Theorem   \ref{lem2} for a $3\times 3$ variable-gain MIMO system.\label{fig:SV_Lya_VG} Dashed lines represent the Lyapunov exponents $\lambda_{\gamma,i}^{(v)}$, \eqref{eq:SV_Lya}; solid lines represent instantaneous realizations of $\frac{1}{n}\log   c_{n,i}^{(v)},$   where, starting from the top, $i=1,2,3.$ For all $i=1,\dots,n$, we set $p_i=n_0=\mu_i = 1.$}
\end{figure}

\begin{figure}
\centering
\includegraphics[scale=0.62]{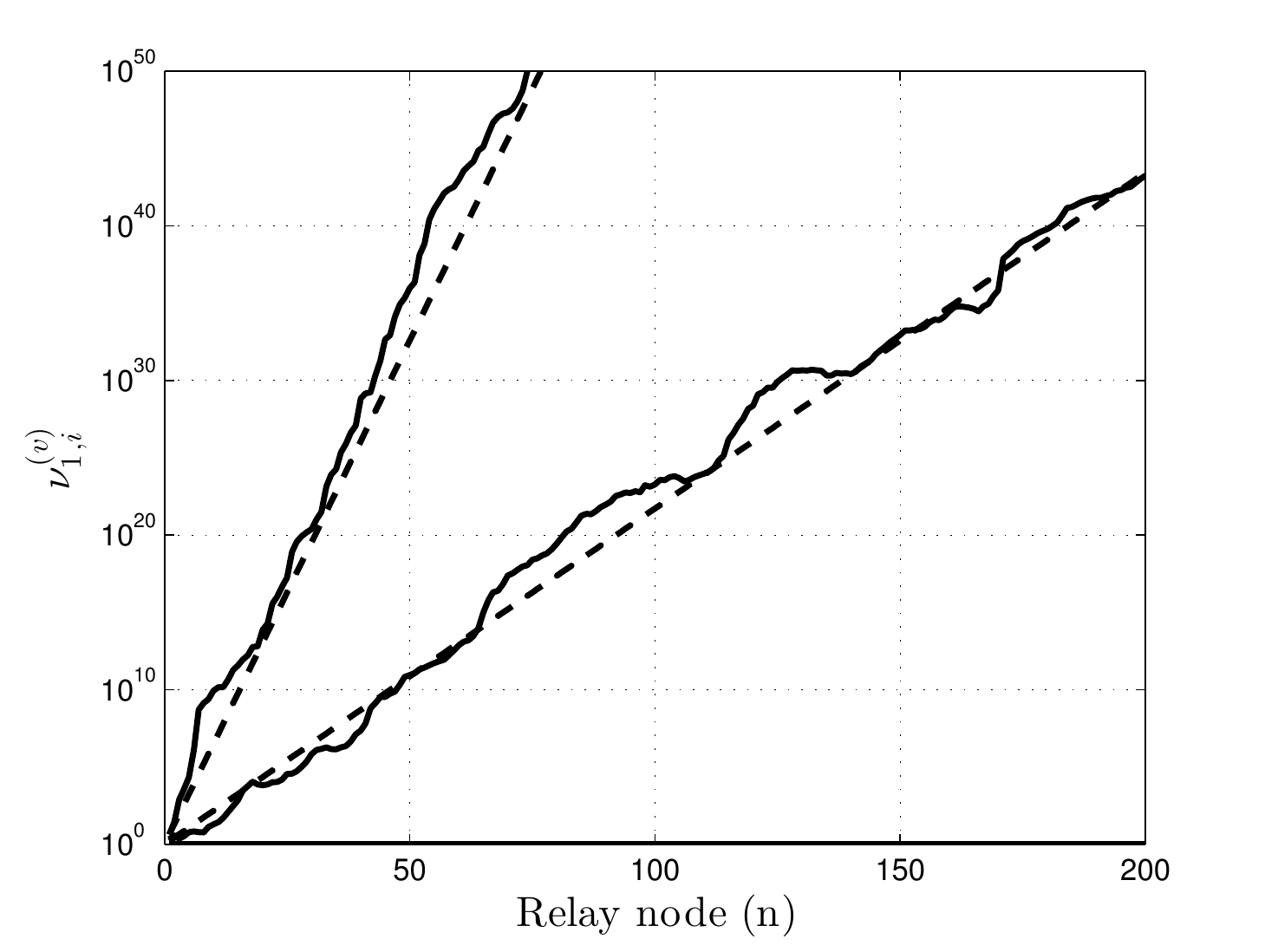}\caption{ \label{fig:realnu_Lya}Figure demonstrating that $\nu_{1,i,n}^{(v)} := c_{n,1}^{(v)}/c_{n,i}^{(v)}$ is given by $\nu_{1,i,n}^{(v)}  =\Theta_{\mathbb P}\left( e^{n\phi_{1,i}^{(v)}} \right)$ for a $3\times 3$ variable-gain MIMO system, see  \eqref{eq:CnumLya}. Dashed lines represent  $\exp\left(n\phi_{1,i}^{(v)}\right)$; solid lines represent instantaneous realizations of $\nu_{1,i,n}^{(v)}$, where, starting from the top, $i=1,\cdots,3.$  For all $i=1,\dots,n$, we set $p_i=n_0=\mu_i = 1.$}
\end{figure}

\begin{figure}
\centering
\includegraphics[scale=0.62]{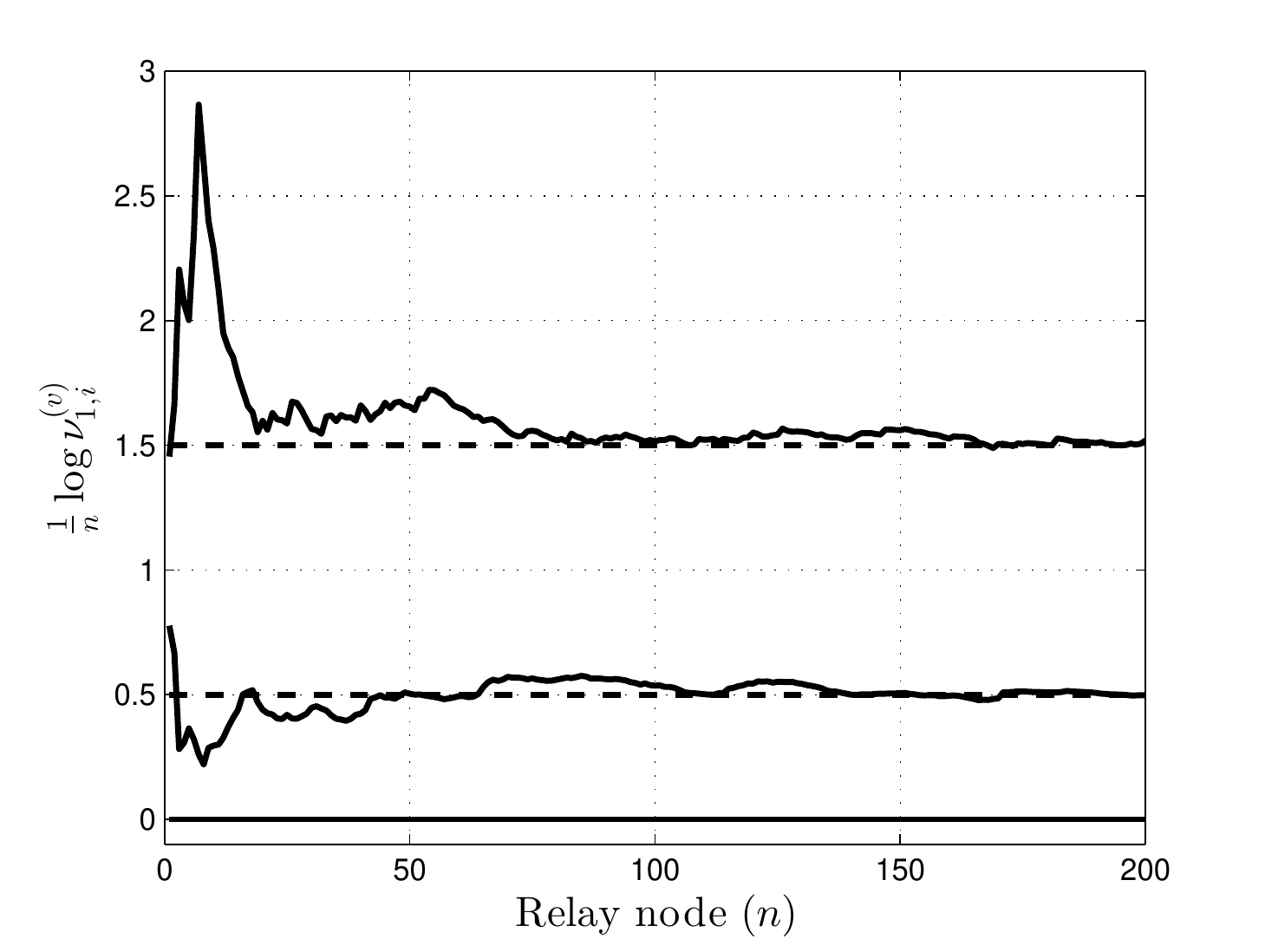}\caption{ \label{fig:nu_Lya}Figure demonstrating \eqref{eq:CnumLya}. Dashed lines represent the Lyapunov difference $\lambda_{\gamma,1}^{(\alpha)}-\lambda_{\gamma,i}^{(\alpha)}$, \eqref{eq:LyaSpread}; solid lines represent instantaneous realizations of $\frac{1}{n}\log  \nu_{1,i,n}^{(v)},$  where, starting from the bottom, $i=1,\dots,4.$  For all $i=1,\dots,n$, we set $p_i=n_0=\mu_i = 1.$}
\end{figure}

Figs. \ref{fig:Cap_FG} and \ref{fig:Cap_Lya} illustrate the second statement of Theorem  \ref{lem2} for a $4\times 4$ fixed-gain  system. The first of these figures shows eigenchannel capacity as a function of network size, and clearly demonstrates that this will trend along a deterministic trajectory governed by the network's Lyapunov exponents. The second of these figures clearly shows convergence in the normalized logarithm of the eigenchannel capacity to the network's Lyapunov exponents.
Figs. \ref{fig:Cap_VG} and \ref{fig:SV_Lya_VG}  illustrate analogous results to those above, but for a variable-gain system. Interestingly, for all these figures we see that convergence to the stated trends occur quickly, sometimes in the order of $5$ to $10$ hops, which attests to the utility of our methods.
Figs. \ref{fig:realnu_Lya} and \ref{fig:nu_Lya} demonstrate \eqref{eq:CnumLya}  as a function of $n$ for a $3\times 3$ variable-gain system. Similar plots occur for fixed-gain. As with above, convergence to the stated trends occurs quickly.


\begin{figure}
\centering
\includegraphics[scale=0.62]{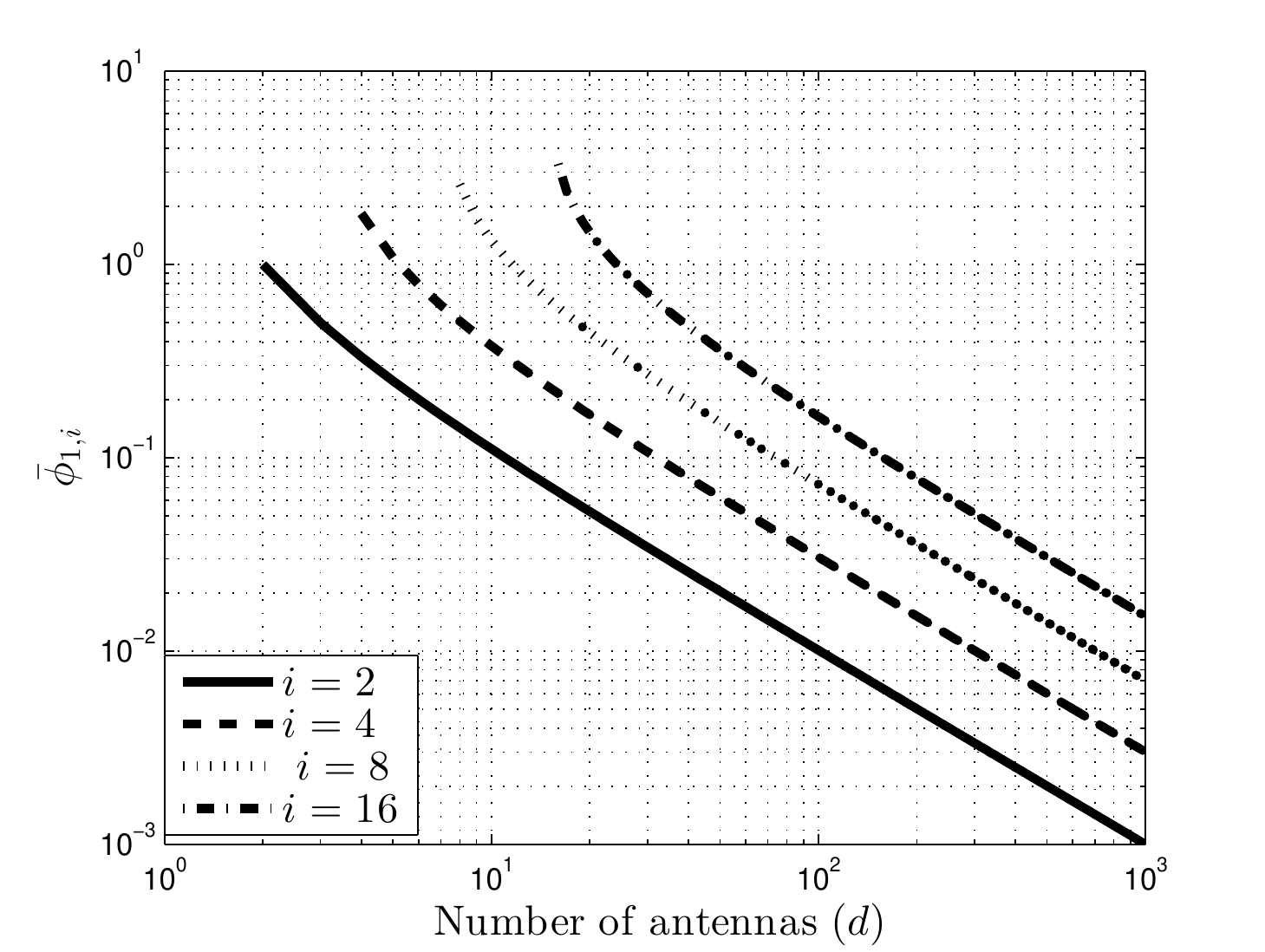}\caption{ \label{fig:phi_d}Figure demonstrating Lemma \ref{lem:1stOrder} for different values of $i$ as a function of the number of antennas.}
\end{figure}

Because of the issues associated with computational complexity (mentioned at the beginning of this section), we were unable to employ Monte Carlo simulations to demonstrate (numerically) the relationship between antenna scaling with respect to number of hops, and the rate at which eigenchannel capacities diverge away from each other (see Lemma \ref{lem:1stOrder}). We do, however, show  Fig. \ref{fig:phi_d}, which plots $ \bar\phi_{1,i}$ as a function of the number of antennas at each node.  In this figure, when $d$ is large the curves are seen to decay linearly on the log-log scale; i.e., they decay like $O(1/d)$ on a linear scale. This observation theoretically illustrates Lemma~\ref{lem:1stOrder} (specifically, \eqref{eq:1orderphi}), and consequently, that if super-linear antenna scaling occurs with respect to the number of hops within the network, the $i$th and $j$th eigenchannel capacities will not exponentially diverge away from each other.

\begin{figure}
\centering
\includegraphics[scale=0.62]{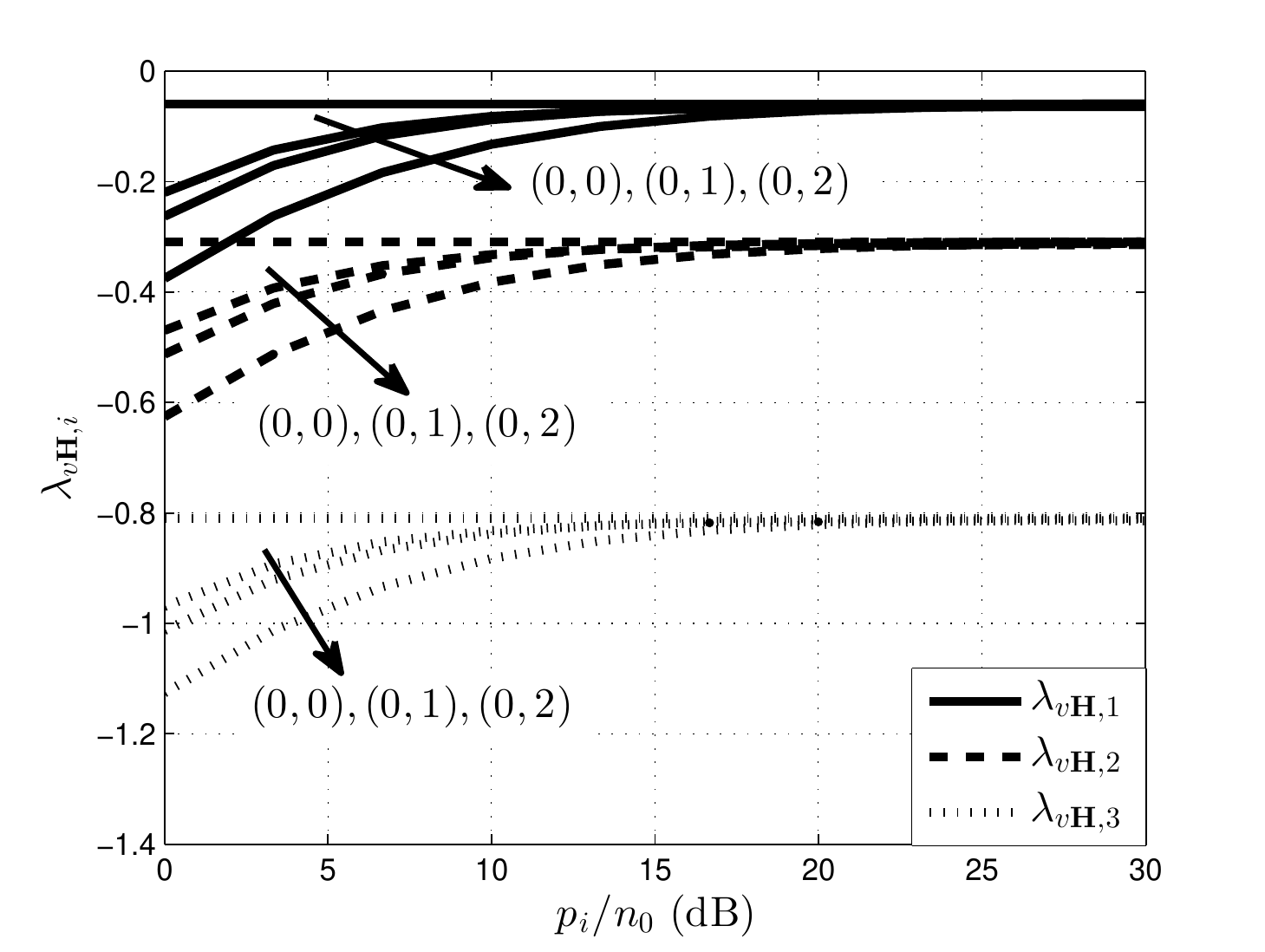}\caption{Figure showing numerically estimated $\lambda_{v\mathbf{H},i}$ (curved lines) for large network ($n=1000$) and its upper bound (straight lines), \eqref{eq:lya_UB}, for a non-homogeneous variable-gain $3\times3$ network. The average channel fading characteristics, $ \mu_i$, are assumed to be log-normally distributed with parameter pairs $(0,1),(0,2)$ and $(0,3)$; i.e., $\mu_i\sim\mathcal{LN}(a,b),\;a=\mathbb E\log \mu_i,\;b= {\mathbb V\log\mu_i},$   . The plot is taken as a function of the normalized transmit power at each node, with $ {p}_i = p_{i-1}$ $\forall \;i$.\label{fig:lya_UB}}
\end{figure}

Finally, Fig. \ref{fig:lya_UB} shows an estimation of $\lambda_{\alpha\mathbf{H},i}^{(v)}$ and its upper bound \eqref{eq:lya_UB} for a variable-gain network as a function of the transmit power at each node for a large network size, $n=1000$. The choice of such a large $n$ is only made to ensure that our results have converged significantly, where smaller values of $n$ may exhibit less smooth plots. For this figure, we assume that the mean channel fading coefficient at the $i$th node is log-normally distributed. It is easy to see that the bound is very tight for large $p_i/n_0$.

 { \section{Conclusion\label{sec:conc}} }
 
In this paper, we have employed the formalism of RDSs to study the scaling properties of the transmit power and end-to-end channel capacity of finite antenna MIMO  AF relay networks.  By employing the RDS formalism, we have been  able to associate Lyapunov exponents (which are classically used to characterize the stability of RDSs) with the MIMO  AF relay network.  Our study has revealed that the exponential growth and/or decay of the transmit power and end-to-end channel capacity  are completely characterized by the network's Lyapunov exponents. Furthermore, our methods can be applied to systems with arbitrary channel fading statistics, provided $\mathbb E\log^+\Vert  \mathbf H_i \Vert<\infty$, where $\mathbf H_i $ is the channel matrix for the $i$th hop; however, in this contribution we focus explicitly on the Rayleigh fading scenario.  We then establish growth laws for the eigenchannel capacity divergence, how this relates to the amplification strategy  and number of antennas at each node, and the cost (in terms of power) associated with multiplexing extra data streams. Finally, we would like to close with the following open question: Can our techniques be extended to study the capacity and power scaling properties of networks employing other (non AF) forwarding strategies?

\appendices



 { \section{\label{apthrm1} Proof of Theorem \ref{lem:TTP}} }

 Firstly,   let 
\begin{equation}
\prod_{i=1}^{n}\mathbf M_i\left[ X_0^T \;\;1\right]^T := \left[ X_n^T\;\;1\right]^T,
\end{equation}
it is easy to see from Definition \ref{def:LE} that $\lambda\left(\mathbf M,\left[ X_0^T \;\;1\right]^T \right)\geq\lambda_{a,1}$. Thus, from Fact \ref{lem:affine}
\begin{equation}\label{eq:Lset}
\lambda\left(\mathbf M,\left[ X_0^T \;\;1\right]^T \right) \in \left\{  \lambda_{\mathbf A,i}\geq \lambda_{a,1} \right\}\cup \lambda_{a,1}=:\mathcal L.
\end{equation}
The proof of the theorem now      follows from Claim \ref{claim1} (mentioned below).
\begin{claim}\label{claim1}
With 
$
\mathcal Y := \{[y_1\;\cdots\;y_d\;1]^T:y_i\in\mathbb C \},
$
the mapping
\begin{equation}
\lambda(\mathbf M, \cdot):\mathcal Y\to \mathcal L\label{eq:surj}
\end{equation}
  is surjective. 
\end{claim}
{\bf{Proof of Claim \ref{claim1}:}}
If $\lambda_{\mathbf A, 1} <\lambda_{a,1}$ then $\mathcal L = \{\lambda_{a,1}\}$, $\lambda\left( \mathbf M,Y \right) = \lambda_{a,1}$ $\forall\;Y\in\mathcal Y$ and the surjectivity of \eqref{eq:surj} is satisfied. Thus, w.l.o.g., we assume that $\exists \;k\leq d$ such that
  \begin{equation}
  \lambda_{\mathbf A,1} >\cdots>\lambda_{\mathbf A,k} \geq \lambda_{a,1}> \lambda_{\mathbf A,k+1}>\cdots >\lambda_{\mathbf A,d}.
  \end{equation}
  In what follows, we consider the scenario in which $\lambda_{\mathbf A,k}>\lambda_{a,1}>\lambda_{\mathbf A,k+1}$. The proof can easily be extended to the case when $  \lambda_{\mathbf A,k}=\lambda_{a,1}$.
  
 Consider the  filtration,
  \begin{equation}
\{0\}=:\mathcal V_{p+1} \subset \mathcal V_p\subset\cdots\subset \mathcal V_1 = \mathbb C^{d+1}\label{eq:Filtration}
\end{equation}
where $Y\in\mathcal  V_i\setminus\mathcal  V_{i+1} \Leftrightarrow \lambda (\mathbf M, Y) = \lambda_i$ (the existence of such a filtration is guaranteed by Fact~\ref{lem:props}.\ref{filtration}).  The proof of Claim \ref{claim1} then follows immediately from Claim \ref{claim2} (mentioned below). 
\begin{claim}\label{claim2}
Let $\mathcal V_i$ be as in \eqref{eq:Filtration} and $\mathcal Y$ be as in Claim \ref{claim1}. Then $(\mathcal V_i \setminus \mathcal V_{i+1})\cap\mathcal Y \neq \emptyset$ for all $i=1,\dots, k+1$, where $\lambda_{\mathbf A,k}>\lambda_{a,1}>\lambda_{\mathbf A,k+1}$.
\end{claim}
 {\bf{Proof of Claim \ref{claim2}:}}
  Claim \ref{claim2} follows immediately from Claim \ref{claim3} (mentioned below).  
\begin{claim}\label{claim3}
Let $\mathcal V_i$ be as in \eqref{eq:Filtration}, $\mathcal Y$ be as in Claim \ref{claim1},  and suppose that $\lambda_{\mathbf{A},k} > \lambda_{a,1} > \lambda_{\mathbf{A},k+1}$. Then:
\newline1)  for all $i\leq k$,   $(\mathcal V_{i}\setminus \mathcal V_{i+1})  \cap\mathcal Y=\emptyset$ implies $ (\mathcal V_{l}\setminus \mathcal V_{l+1})  \cap\mathcal Y = \emptyset$ for all $l<i$,
\newline 2)  $(\mathcal V_{1}\setminus \mathcal V_{2})  \cap\mathcal Y\neq \emptyset$.
\end{claim}
 {\bf{Proof of Claim \ref{claim3}:}}
We will begin by proving the first part of the claim. To do this, we first note the following: all the Lyapunov exponents have multiplicity $1$ (i.e., they are distinct); consequently, from Fact \ref{filtration}, $\dim \mathcal  V_j-\dim \mathcal V_{j+1} = 1$ $\forall\;j$ and
\begin{equation}
\dim \mathcal V_j =d+2-j.\label{eq:dim}
\end{equation}

 Clearly, 
\begin{equation}
(\mathcal V_{i}\setminus\mathcal  V_{i+1})  \cap\mathcal Y = \emptyset\Leftrightarrow \mathcal V_{i}\cap \mathcal Y = \emptyset\;\mathrm{or}\;\mathcal Y\subseteq \mathcal V_{i+1}\subset\cdots\subset \mathcal V_1.\label{eq:LRimp}
\end{equation}
However, if $\mathcal V_i\cap \mathcal Y=\emptyset$ is satisfied, it can be seen that  because $\mathcal V_{i}$ is a vector space all vectors in $\mathcal V_{i}$ must have their $(d+1)$th element equal to zero. Thus,
\begin{eqnarray}
&&\!\!\!\!\!\!\!\!\!\!\!\!\mathcal V_{i} \cap \mathcal Y = \emptyset \nonumber\\
&\Rightarrow& \!\!\!\!\mathcal V_{i} = \left\{X=   [y_1 \; \cdots \;y_d\;0]^T : \; \lambda(\mathbf M, X) \leq \lambda_{\mathbf A , i} \right\}\nonumber\\
&\Rightarrow &\!\!\!\!\dim \mathcal V_i  =  \dim\left\{X'=   [y_1 \; \cdots \;y_d]^T : \; \lambda( \mathbf A, X') \leq \lambda_{\mathbf A , i} \right\}\nonumber\\
 &&\quad \quad\quad =   d +1 - i .
\end{eqnarray}
  But from \eqref{eq:dim}, $\dim   \mathcal V_{i}=d+2-i$, so $  \mathcal V_{i} \cap \mathcal Y= \emptyset$ gives us a contradiction, \eqref{eq:LRimp} becomes
  \begin{eqnarray}
  (\mathcal V_{i}\setminus \mathcal V_{i+1})\cap \mathcal Y = \emptyset&\Leftrightarrow& \mathcal Y \subseteq \mathcal V_{i+1}\subset\cdots\subset\mathcal  V_1, \label{eq:LRimp2}
  \end{eqnarray}
  and
   \begin{eqnarray}
  (\mathcal V_j\setminus\mathcal  V_{j+1})\cap\mathcal Y =\emptyset,\;\forall \;j\leq i.\label{eq:Rimp2}
  \end{eqnarray}
This proves the first part of the Claim.

We will now prove the second part of the claim. From \eqref{eq:LRimp2} we have $(\mathcal V_1\setminus\mathcal  V_2) \cap \mathcal Y =\emptyset \Leftrightarrow \mathcal V_2\supseteq \mathcal Y$. But $\mathcal Y$ contains a $d$ dimensional subspace $\mathcal A := \{  [y_1 \; \cdots \;y_d\;0]^T : y_i \in \mathbb C\}$, and $\mathcal V_2$ is also $d$ dimensional, so
\begin{equation}
\mathcal V_2 \supseteq\mathcal Y\supseteq \mathcal A\Rightarrow\mathcal A =\mathcal  V_2\Rightarrow \mathcal A = \mathcal Y .\label{eq:contrad}
\end{equation}
  But $\mathcal A\subset \mathcal Y$, so from \eqref{eq:contrad} $\mathcal V_2 \supseteq\mathcal Y$ gives us a contradiction. Thus $\mathcal  V_2 \not\supseteq \mathcal Y$, which (from \eqref{eq:LRimp2}) gives 
  \begin{equation}
  (\mathcal V_1\setminus \mathcal V_2)\cap \mathcal Y \neq \emptyset . \label{eq:almost}
  \end{equation}
  This completes the proof.

 { \section{\label{apAA} Proof of Lemma \ref{lem:AlwaysPos}} }
We have
\begin{eqnarray}
 \lim_{n\to\infty}\frac{1}{n}\log||X_n || &= &\!\!\lim_{n\to\infty}\frac{1}{n}\log||\mathbf A_n X_{n-1}  - R_n +2R_n || \nonumber\\
\!\!\!\!&\leq&\!\! \max \left\{\lim_{n\to\infty}\frac{1}{n}\log||\mathbf A_n X_{n-1}  - R_n||  ,\right.\nonumber\\
& &\qquad\qquad\qquad\left.\lim_{n\to\infty}\frac{1}{n}\log||2R_n||     \right\}\nonumber  \\
\!\!\!\!&\overset{a.s.}{=}&\!\!\max \left\{\lim_{n\to\infty}\frac{1}{n}\log||X_n ||  , 0  \right\} \label{eq:SEproof},
 \end{eqnarray}
where the second line follows from Lemma  \ref{lem:holdwitheq} (below) and the  last line follows from the symmetry of $R_n$ and that $R_n\overset{a.s.}{\neq} 0$.
If  $$\lim_{n\to\infty}\frac{1}{n}\log||X_n || \geq 0,$$  our result is reached trivially; if $$\lim_{n\to\infty}\frac{1}{n}\log||X_n ||  = \lambda< 0,$$  from  Lemma  \ref{lem:holdwitheq} (below),  the line above \eqref{eq:SEproof}  holds with equality, which gives $\lambda=0$. This contradicts our assumption that $\lambda < 0$. Therefore, $$\lim_{n\to\infty}\frac{1}{n}\log||X_n || \geq 0.$$
This completes the proof.

  \begin{lemma}\label{lem:holdwitheq}
For $\alpha_n,\beta_n\in\mathbb C^d$,
\begin{multline}
\lim_{n\to\infty}\frac{1}{n}\log||\alpha_n+\beta_n||\\ \leq \max\left\{ \lim_{n\to\infty}\frac{1}{n}\log||\alpha_n || \;, \;\lim_{n\to\infty}\frac{1}{n}\log||\beta_n|| \right\},\label{eq:eqineq}
\end{multline}
where equality holds when   
\begin{equation}
  \lim_{n\to\infty}\frac{1}{n}\log||\alpha_n ||  \neq \lim_{n\to\infty}\frac{1}{n}\log||\beta_n||    . 
\end{equation}
\end{lemma}
\begin{IEEEproof}
For $\alpha_n,\beta_n\in\mathbb C^d$,
\begin{multline}
\lim_{n\to\infty}\frac{1}{n}\log||\alpha_n+\beta_n|| \\ \leq \max\left\{ \lim_{n\to\infty}\frac{1}{n}\log||\alpha_n || \;, \;\lim_{n\to\infty}\frac{1}{n}\log||\beta_n|| \right\}\label{eq:eqineq}
\end{multline}
since $||\alpha_n +\beta_n||\leq 2\max\{  ||\alpha_n||, ||\beta_n ||\} $. To show that \eqref{eq:eqineq}   holds with equality when
\begin{equation}
 \lim_{n\to\infty}\frac{1}{n}\log||\alpha_n ||  \neq  \lim_{n\to\infty}\frac{1}{n}\log||\beta_n|| ,
\end{equation}
 w.l.o.g., we assume that $\lim_{n\to\infty}\frac{1}{n}\log ||\alpha_n||<\lim_{n\to\infty}\frac{1}{n}\log ||\beta_n||$. Eq. \eqref{eq:eqineq} then gives us
\begin{eqnarray}
\lim_{n\to\infty}\frac{1}{n}\log ||\alpha_n+\beta_n||&\leq& \lim_{n\to\infty}\frac{1}{n}\log ||\beta_n||\label{eq:logbetan}\\
 &\leq& \lim_{n\to\infty}\frac{1}{n}\log ||\alpha_n+\beta_n - \alpha_n|| \nonumber\\
\!\!\!\!\!\!\!\!\!&\leq& \max \left\{  \lim_{n\to\infty}\frac{1}{n}\log ||\alpha_n+\beta_n||,\right.\nonumber\\
&&\;\;\;\;\qquad\left.\lim_{n\to\infty}\frac{1}{n}\log ||\alpha_n||  \right\}\label{eq:lomaxalphagbetan}
\end{eqnarray}
It follows that if $\lim_{n\to\infty}({1}/{n})\log ||\alpha_n+\beta_n|| < \lim_{n\to\infty}\frac{1}{n}\log ||\alpha_n||$ then (from \eqref{eq:logbetan} and \eqref{eq:lomaxalphagbetan}) $\lim_{n\to\infty}({1}/{n})\log ||\beta_n|| \leq \lim_{n\to\infty}({1}/{n})\log ||\alpha_n||  $ which contradicts our
assumption. Consequently, $\lim_{n\to\infty}({1}/{n})\log ||\beta_n||$  is sandwiched either side by   $ \lim_{n\to\infty}({1}/{n})\log ||\alpha_n+\beta_n||  $ and so must be equal to it.
\end{IEEEproof}


 {\section{\label{apB}Proof of Theorem \ref{lem2}}}

 Theorem \ref{lem2} contains two statements. We prove these separately in the following two subsections. 

\subsection{ {First Statement}}
We prove the first statement in two parts.  Each of these parts will involve manipulating the inverse of $\left( \mathbf  R_{\mathcal{I},n}^{(\alpha)} \mathbf R_{\mathcal{N},n}^{(\alpha)-1} \right) $, which is given by
\begin{eqnarray}
\left(\mathbf  R_{\mathcal{I},n}^{(\alpha)}\mathbf  R_{\mathcal{N},n}^{(\alpha)-1} \right)^{-1}  \!\!\!\!\!\!\!\!& =& \!\!\!\!\mathbf  R_{\mathcal{N},n}\mathbf R_{\mathcal{I},n}^{(\alpha)-1}\nonumber\\
\!\!\!\!\!\!\!\!\!\!\!\!& = &  \!\!\!\! \left(\mathbf  R_{\mathcal{I},n}^{(\alpha)-1} + \right.\label{eq:Time_Rev1} 
 \\ &&\left.\sum_{l=2}^{n}\mathbf H_n\cdots \mathbf H_l\mathbf  R_{\mathcal{I},l-1}^{(\alpha)-1} \mathbf H_l^{-1}\cdots \mathbf H_n^{-1} \right), \nonumber
\end{eqnarray} 
where, without loss of generality, we have assumed that $n_0 =~1$. 
The first part constructs an upper bound on the limit in question. The second part constructs a lower bound  on the same limit, which is identical to the lower bound. This proves the first part of the theorem.
 \subsubsection{ {Upper Bound} }

Our aim is to show that 
\begin{equation}
\lim_{n\to\infty}\frac{1}{n}\log\mathcal E_i\left(\mathbf R_{\mathcal{I},n}^{(\alpha)}\mathbf R_{\mathcal{N},n}^{(\alpha)-1} \right) \overset{a.s.}{\leq} \min\{0, 2\lambda_{\alpha\mathbf{H},i}  \}.\label{eq:thrm2UB}
\end{equation}
By noting that 
$$\mathcal E_i \left(    \mathbf  R_{\mathcal{I},1}^{(\alpha)-1}   \right) \overset{a.s.}{\neq} 0,$$
and does not depend on $n$, we obtain
\begin{equation}\lim_{n\to\infty}\frac{1}{n}\log\mathcal E_i \left(    \mathbf  R_{\mathcal{I},1}^{(\alpha)-1}   \right) \overset{a.s.}{=} 0.\label{eq:eq:as0}\end{equation}
Also, from the definition of $\mathbf  R_{\mathcal{I},n}^{(\alpha)}$ and property \ref{itemSV} of Fact~\ref{lem:props}, it is clear that 
\begin{equation}
\lim_{n\to\infty}\frac{1}{n}\log\mathcal E_i \left(    \mathbf  R_{\mathcal{I},n}^{(\alpha)-1}   \right)  \overset{a.s.}{=} -2\lambda_{\alpha\mathbf H, i}.\label{eq:eq:aslambda}
\end{equation}
By combining \eqref{eq:eq:as0}, \eqref{eq:eq:aslambda}, and 
\begin{equation}
 - \max\{0, -a\} = \min\{0,a\},\;a\in\mathbb R,\label{eq:identity}
\end{equation}
the upper bound of \eqref{eq:thrm2UB} follows immediately from  the following claim (the proof of which is given at the end of Appendix~\ref{apB}).
\begin{claim}\label{claim4}
The eigenvalues of $ \left(\mathbf  R_{\mathcal{I},n}^{(\alpha)}\mathbf  R_{\mathcal{N},n}^{(\alpha)-1} \right)^{-1}$ are bound in the following way:
\begin{multline}
\mathcal E_i \left( \left(\mathbf  R_{\mathcal{I},n}^{(\alpha)}\mathbf  R_{\mathcal{N},n}^{(\alpha)-1} \right)^{-1}\right)\\ \geq \max\left\{ \mathcal E_i \left(  \mathbf  R_{\mathcal{I},n}^{(\alpha)-1}\right)    ,    \mathcal E_i \left(  \mathbf  R_{\mathcal{I},1}^{(\alpha)-1}\right)        \right\},\label{eq:maxevalues}
\end{multline}
\end{claim}

 \subsubsection{ {Lower Bound}}
We will now provide the second part of the proof (constructing the lower bound). 
 To begin, let us   introduce the following  RDS\footnote{Note, \eqref{eq:Y} is a backward RDS as per \eqref{eq:backwardRDS}.}, which will be exploited in a moment:
\begin{equation}
Y_n: = {\mathbf{M}}_{1}^{(\alpha)}\cdots {\mathbf{M}}_{n}^{(\alpha)}\left[\begin{array}{c}
\hat{Y}_{0}\\
1
\end{array}\right]\label{eq:Y},
\end{equation}
where 
\begin{equation}
 {\mathbf{ M}}_{i}^{(\alpha)} :=  \left[\begin{array}{cc}
\frac{1}{\alpha_{i}}\mathbf{H}_{i}^{-1}   &  \hat Y_{0}\\
\bf{0}^{T} &\pm 1
\end{array}\right]\label{eq:hatM_Matrix}.
\end{equation}
To allow us to describe the mechanism by which the sign of $\pm 1$ is chosen in the bottom right corner of \eqref{eq:hatM_Matrix}, we must first establish the inner product of $Y_n$.
The inner product of $Y_n$ is given by
\begin{align}
\Vert Y_n\Vert^2  =& \overset{\mathrm{First \; inner \;product\; term}}{\overbrace{\hat{Y}_0^{\dagger}\left(\overline{\mathbf{R}}_{\mathcal{I},n}^{(\alpha)-1}+\sum_{l=1}^{n-1}\overline{\mathbf{R}}_{\mathcal{I},l}^{(\alpha)-1} +   \mathbf I_d \right)^\dagger\hat{Y}_0}}\nonumber\\
 &\!\!\!\!\!\!\!\!\!\!\!\!\!\!\overset{\mathrm{Second \; inner \;product\; term}}{\overbrace{ \pm  \hat{Y}_0^{\dagger}\left(\frac{1}{\alpha_{j}  }\left(\mathbf{H}_{n}^{-1}\right)^{\dagger} \cdots  \left(\mathbf{H}_{1}^{-1}\right)^{\dagger}\mathbf{H}_{1}^{-1}\cdots \mathbf{H}_{n-1}^{-1}  \right. \prod_{j =1}^{n}\frac{1}{\alpha_{j}  }\prod_{j = 1}^{n-1}  }}\nonumber\\
&\pm\cdots\pm    \left(\mathbf{H}_{n}^{-1}\right)^{\dagger} \cdots  \left(\mathbf{H}_{1}^{-1}\right)^{\dagger}\mathbf{H}_{1}^{-1} \frac{1}{\alpha_{1}  } \prod_{j =1}^{n}\frac{1}{\alpha_{j}  } \nonumber \\
&\left. \qquad\quad\pm\cdots\pm  \left(\mathbf{H}_{1}^{-1}\right)^{\dagger} \frac{1}{\alpha_{1}  }  \right)\hat{Y}_0, \label{eq:innerproductYn}
\end{align}
where
\begin{equation}
\overline{\mathbf R}_{\mathcal I,i}^{(\alpha)-1} :=  \mathbf{H}_{i}^{-1} \cdots  \mathbf{H}_{1}^{-1}\left(\mathbf{H}_{1}^{-1}\right)^{\dagger}  \cdots  \left(\mathbf{H}_{i}^{-1}\right)^{\dagger}\prod_{j = 1}^{i}\frac{1}{\alpha_{j}^2 }.
\end{equation}
It is clear that the second inner product term  is a real number that, at the moment,   may be either positive or negative. However, there is nothing stopping us from ensuring that this is  strictly positive by appropriately selecting the sign of $\pm 1$ in \eqref{eq:hatM_Matrix}; for, the RDS is permitted to remember the past, and predict the future \cite{arnold1998random}. This is the mechanism that we will use to select the sign. Furthermore,  performing sign selection in this way will not affect the Lyapunov exponents of the system in question (Fact~\ref{lem:affine}). We now have the following upper bound on   the first inner product term of \eqref{eq:innerproductYn}, which will be exploited later on:
\begin{equation}
\Vert Y_n\Vert^2 \geq   { {\hat{Y}_0^{\dagger}\left(\overline{\mathbf{R}}_{\mathcal{I},n}^{(\alpha)-1}+\sum_{l=1}^{n-1}\overline{\mathbf{R}}_{\mathcal{I},l}^{(\alpha)-1} + \mathbf I_d \right)^\dagger\hat{Y}_0}}.\label{Yn:bound}
\end{equation}

 It can already be seen that \eqref{eq:Time_Rev1} is remarkably similar to the first inner product term of \eqref{eq:innerproductYn}. We will now show that this similarity is not superficial, and that
\begin{multline}
\lim_{n\to\infty}\frac{1}{n}\log\mathcal E_i\left(\left(\mathbf R_{\mathcal{I},n}^{(\alpha)}\mathbf R_{\mathcal{N},n}^{(\alpha)-1} \right)^{-1}\right)  \\\leq \lim_{n\to\infty}\frac{1}{n}\log\mathcal E_i\left(\overline{\mathbf{R}}_{\mathcal{I},n}^{(\alpha)-1}+\sum_{l=1}^{n-1}\overline{\mathbf{R}}_{\mathcal{I},l}^{(\alpha)-1} +  \mathbf I_d \right).
\label{eq:lims12}
\end{multline} 
 To do this, note that  
\begin{multline}
\!\!\!\!\!\!\!\!\!\!\mathcal E_i\left\{\mathbf H_n\cdots \mathbf H_l\mathbf  R_{\mathcal{I},l-1}^{(\alpha)-1} \mathbf H_l^{-1}\cdots \mathbf H_n^{-1} \right\}  \\= \mathcal E_i\left\{\mathbf R_{\mathcal{I},l-1}^{(\alpha)-1}\right\} 
 \overset{d}{=}   \mathcal E_i\left\{\overline{\mathbf R}_{\mathcal{I},l-1}^{(\alpha)-1}\right\} .\label{eq:eqdist11}
\end{multline}
 Consequently,
\begin{align}
 \mathcal E_i\left( \left(\mathbf  R_{\mathcal{I},n}^{(\alpha)} \mathbf R_{\mathcal{N},n}^{(\alpha)-1} \right)^{-1}\right)  & \overset{d}{=}  \mathcal E_i\left( {\mathbf{R}}_{\mathcal{I},n}^{(\alpha)-1}+\sum_{l=1}^{n-1}{\mathbf{R}}_{\mathcal{I},l}^{(\alpha)-1}\right) \label{eq:eqdist1} \\ &\overset{d}{=}  
  \mathcal E_i\left( \overline{\mathbf{R}}_{\mathcal{I},n,1}^{(\alpha)-1}+\sum_{l=1}^{n-1}\overline{\mathbf R}_{\mathcal{I},l}^{(\alpha)-1}\right)   \label{eq:eqdist1234} \\
&\leq  \mathcal E_i\left(\overline{\mathbf{R}}_{\mathcal{I},n,1}^{(\alpha)-1}+\sum_{l=1}^{n-1}\overline{\mathbf R}_{\mathcal{I},l}^{(\alpha)-1} + \mathbf{I}_d\right) , \label{eq:eqdist}
\end{align}
where \eqref{eq:eqdist1} follows from the first equality of \eqref{eq:eqdist11}, \eqref{eq:eqdist1234} follows from  the  second equality of \eqref{eq:eqdist11}, and \eqref{eq:eqdist} follows trivially from \eqref{eq:eqdist1234}.
 
 With \eqref{eq:eqdist1} and \eqref{eq:eqdist}, we have  shown \eqref{eq:lims12}. 
The right hand side of \eqref{Yn:bound} is known to be equal to the $i$th eigenvalue when $\hat Y_0$ is an $i$th unit eigenvector. Combining this fact with \eqref{eq:eqdist} gives us
\begin{equation}
\lim_{n\to\infty}\frac{1}{n}\log\mathcal E_i\left(\left(\mathbf R_{\mathcal{I},n}^{(\alpha)}\mathbf R_{\mathcal{N},n}^{(\alpha)-1} \right)^{-1}\right)  \leq \lim_{n\to\infty}\frac{1}{n}\log \Vert Y_n\Vert^2 .
\label{eq:lims13}
\end{equation} 
  But the limit  on  the right hand side of \eqref{eq:lims13}, when $Y_n$ is given by \eqref{eq:Y}, is given by  Theorem \ref{lem:TTP}. Thus, \eqref{eq:lims13} and  Theorem~\ref{lem:TTP} give  us
\begin{equation}\lim_{n\to\infty} \frac{1}{n}\log \mathcal E_i\left(\mathbf R_{\mathcal{I},n}^{(\alpha)}\mathbf  R_{\mathcal{N},n}^{(\alpha)-1} \right) \geq - \max\left\{-2\lambda_{ \alpha\mathbf{H} ,i},0\right\},\end{equation}
which can then be combined with \eqref{eq:identity} to yield the lower bound.



 {
\subsection{Second Statement\label{Corollary}}}

From the first statement of Theorem \ref{lem2}, we have
\begin{multline}
\mathbb{P}\left[\log\left(e^{n\lambda_{\gamma,i}^{(\alpha)}-o\left(n\right)}\!+\!1\right)\leq c_{n,i}^{(\alpha)}\leq\log\left(e^{n\lambda_{\gamma,i}^{(\alpha)}+o\left(n\right)}\!+\!1\right)\right]\\\to  1,\nonumber
\end{multline}
which gives
\begin{align*}
\!\mathbb{P}\!\left[e^{n\lambda_{\gamma,i}^{(\alpha)}\!-\! o\left(n\right)} +  O\left(e^{2n\lambda_{\gamma,i}^{(\alpha)}}\right)\leq\right.\qquad\qquad\qquad\qquad&\\
\left.\! c_{n}^{(\alpha)}\!\leq\! e^{n\lambda_{\gamma,i}^{(\alpha)}\!+\! o\left(n\right)} +  O\left(e^{2n\lambda_{\gamma,i}^{(\alpha)}}\right)\right] & \to  1\\
\Rightarrow\mathbb{P}\left[e^{n\lambda_{\gamma,i}^{(\alpha)}-o\left(n\right)}O\left(1\right)\leq c_{n}^{(\alpha)}\leq e^{n\lambda_{\gamma,i}^{(\alpha)}+o\left(n\right)}O\left(1\right)\right]&\to 1\\
\Rightarrow\mathbb{P}\left[e^{n\lambda_{\gamma,i}^{(\alpha)}-o\left(n\right)}\leq c_{n}^{(\alpha)}\leq e^{n\lambda_{\gamma,i}^{(\alpha)}+o\left(n\right)}\right]&\to 1
\end{align*}
where the first line follows from the Taylor expansion of $\log(1+x)$ about $x=0$ and the second line follows by factoring $e^{n\lambda_{\gamma}^{(\alpha)}\pm o(n)}$ from the left and right sides of the second line, respectively, and noting that $\lambda_{\gamma,i}^{(\alpha)} \leq 0$.

{\bf{Proof of Claim \ref{claim4}}:}
An immediate consequence of the dual Lidskii inequality \cite{tao2012topics} is that
 \begin{equation}
 \mathcal{E}_i\left( \mathbf A  +  \mathbf B  \right) \geq \mathcal E_i\left(\mathbf A\right)  + \mathcal E_d\left(\mathbf B\right),\label{eq:lid}
 \end{equation}
 which applies to $d\times d$ Hermitian matrices $\mathbf{A}$ and $\mathbf{B}$. Combining \eqref{eq:lid} with the fact that the summands in \eqref{eq:Time_Rev1} are positive definite (i.e., they have positive eigenvalues), gives us
 \begin{align}
\mathcal E_i \left( \left(\mathbf  R_{\mathcal{I},n}^{(\alpha)}\mathbf  R_{\mathcal{N},n}^{(\alpha)-1} \right)^{-1}\right)& \geq  \mathcal E_i \left( \mathbf  R_{\mathcal{I},n}^{(\alpha)-1}\right)\quad\mathrm{and}\\ 
\mathcal E_i \left( \left(\mathbf  R_{\mathcal{I},n}^{(\alpha)}\mathbf  R_{\mathcal{N},n}^{(\alpha)-1} \right)^{-1}\right) &\geq \mathcal E_i \left( \mathbf H_n\cdots \mathbf H_2\mathbf  R_{\mathcal{I},1}^{(\alpha)-1}\right. \\&\quad\quad\quad\quad\left.\times\mathbf H_2^{-1}\cdots \mathbf H_n^{-1}\right).\label{eq:doublebound}
\end{align}
Claim \ref{claim4} follows immediately from \eqref{eq:doublebound} after noting that 
$$ \mathcal E_i \left( \mathbf H_n\cdots \mathbf H_2\mathbf  R_{\mathcal{I},1}^{(\alpha)-1} \mathbf H_2^{-1}\cdots \mathbf H_n^{-1}\right) = \mathcal E_i \left(  \mathbf  R_{\mathcal{I},1}^{(\alpha)-1} \right).$$

 {
\section{\label{ap:AVG_Grow}Proof of Lemma \ref{lem:AVGPWR}}}

Lemma \ref{lem:AVGPWR} contains two statements.  We prove these separately in the following two subsections. 

\subsection{First Statement}

The Lyapunov exponents of the matrix product that describes the progression of $\mathcal I_n$, \eqref{eq:SISO_Infor_Form}, follow immediately from \cite[Proposition 1]{Lyap_CG}. The Lyapunov exponents of $X_n^{(\alpha)}$, \eqref{eq:SISO_Infor_Form}, then follow immediately from Theorem \ref{lem:TTP}. Combining these with  \eqref{lem:traj}, we obtain  \eqref{eq:Itraj} and \eqref{eq:Xtraj}.

\subsection{Second Statement}

For the second statement, we begin by showing that the limit is greater than or equal to zero for both fixed-gain and  variable-gain:
\begin{equation}
\lim_{n\to\infty} \frac{1}{n} \log \frac{p_n}{p_0} \geq \lim_{n\to\infty} \frac{1}{n} \log \frac{\alpha_n^2n_0}{p_0} \overset{a.s.}{=}  0 , \label{eq:gt0}
\end{equation}
where the \emph{almost sure} equality becomes an equality for fixed-gain.
 For fixed-gain, the stated result then follows immediately from \eqref{eq:gt0} and \eqref{eq:LIFGB}.
For  variable-gain, the stated result then follows immediately from \eqref{eq:gt0} and \eqref{eq:lya_UB}.

 {
\section{\label{spreadUB}Proof of Lemma \ref{lem9}}}

The lower bound follows trivially from \eqref{eq:SV_Lya} and \eqref{eq:LyaSpread}. By noting that $\lambda_{\alpha\mathbf{H},1}^{(\alpha)} > \lambda_{\alpha\mathbf{H},i}^{(\alpha)}\geq 0 \Leftrightarrow \lambda_{\gamma,1}^{(\alpha)} = \lambda_{\gamma,i}^{(\alpha)}= 0$, we obtain equality of the bound.
For the upper bound, we need to prove that
\begin{equation}
a - b \geq \min\{0,a\} - \min\{0,b\}
\end{equation}
for $ a\geq b$. To do this, we need to check the following three cases:
\begin{enumerate}
\item $a\geq 0,\;b\geq 0$, $a\geq b$;
\item $a\geq 0,\;b\leq0$;
\item$a\leq0,\;b\leq0$, $a\geq b$;
\end{enumerate}
which can be done trivially. Equality of the upper bound occurs when $b\leq a\leq 0 $. Finally, to obtain the if and only if statements, we need to show that $a> 0$ and $b< 0$ implies that 
\begin{equation}
a-b>\min\{0,a\}-\min\{0,b\} > 0,
\end{equation}
which can be done trivially. The independence of fixed-gain or variable-gain implementation is trivial.

 { \section{\label{ap2}}}

 \begin{lemma}\label{lem:UB}
For the fixed-gain network,
 \begin{equation}
 \lambda_{f \mathbf{H},j}   \leq   \frac{1}{2}\left(  \lim_{n\to\infty} \frac{1}{n}\log\frac{p_n}{p_0}-\log d + \psi(d-j+1)\right)\label{eq:LIFGB},
 \end{equation}
For the variable-gain network, 
\begin{multline}
\lambda_{v\mathbf{H},i}  \leq    \frac{1}{2}  \left(\lim_{n\to\infty} \frac{1}{n}\log\frac{p_n}{p_0} + \log(d) - \psi\left( d^2 \right) \right.\\\left.+ \psi\left( d - i+1 \right) \right) .
 \label{eq:lya_UB}
\end{multline}
\end{lemma}
 \begin{IEEEproof}
The first equation, \eqref{eq:LIFGB}, is obtained from \eqref{eq:LI} by noting that $$L(f^2\mu) \leq  \lim_{n\to\infty} \frac{1}{n}\log\frac{p_n}{p_0}-\log d.$$
For the second equation,  \eqref{eq:lya_UB}, we have
\begin{align}
L(v^2\mu) = &\lim_{n\to\infty}\frac{1}{n}\sum_{i=1}^n \log\left( \frac{ {p}_i}{\frac{p_{i-1}}{d} \Vert\mathbf  H_i\Vert_F^2+dn_0}\mu_i\right)\nonumber\\
 \leq &\lim_{n\to\infty}\frac{1}{n}\sum_{i=1}^n \log\left( \frac{d  {p}_i}{p_{i-1} \Vert\mathbf  H_i\Vert_F^2 }\mu_i\right)\nonumber\\
 =&\lim_{n\to\infty}\frac{1}{n} \log \frac{ {p}_n}{p_{0}}  +\log (d) - \psi(d^2).
\end{align}
where the final  line follows from $\mathbb E  \log \left\Vert\mathbf  H_i \right\Vert_F^2/\mu_i = \psi(d^2)$. From \eqref{eq:LI}, the stated result follows immediately.
\end{IEEEproof}

 \bibliographystyle{ieeetr}
\bibliography{skel_v8}
\begin{IEEEbiography}
    [{\includegraphics[width=1in,height=1.25in,clip,keepaspectratio]{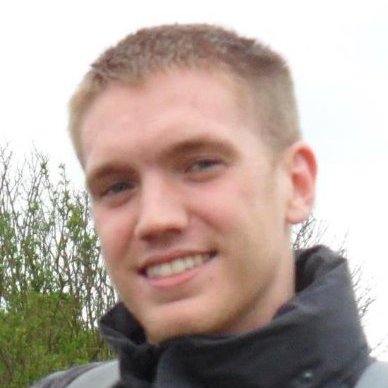}}]{Mr David E. Simmons} graduated from the University of Central Lancashire with a degree in Mathematics (2011). He then went on to study for an MSc in Communications Engineering at the University of Bristol (2012). He is currently a DPhil student at the University of Oxford, where his research has focused predominantly on amplify-and-forward relay networks.
 During his DPhil, David was the recipient of a `best paper' award. David's research interests include information theory and communication theory.
\end{IEEEbiography}
\begin{IEEEbiography}
    [{\includegraphics[width=1in,height=1.25in,clip,keepaspectratio]{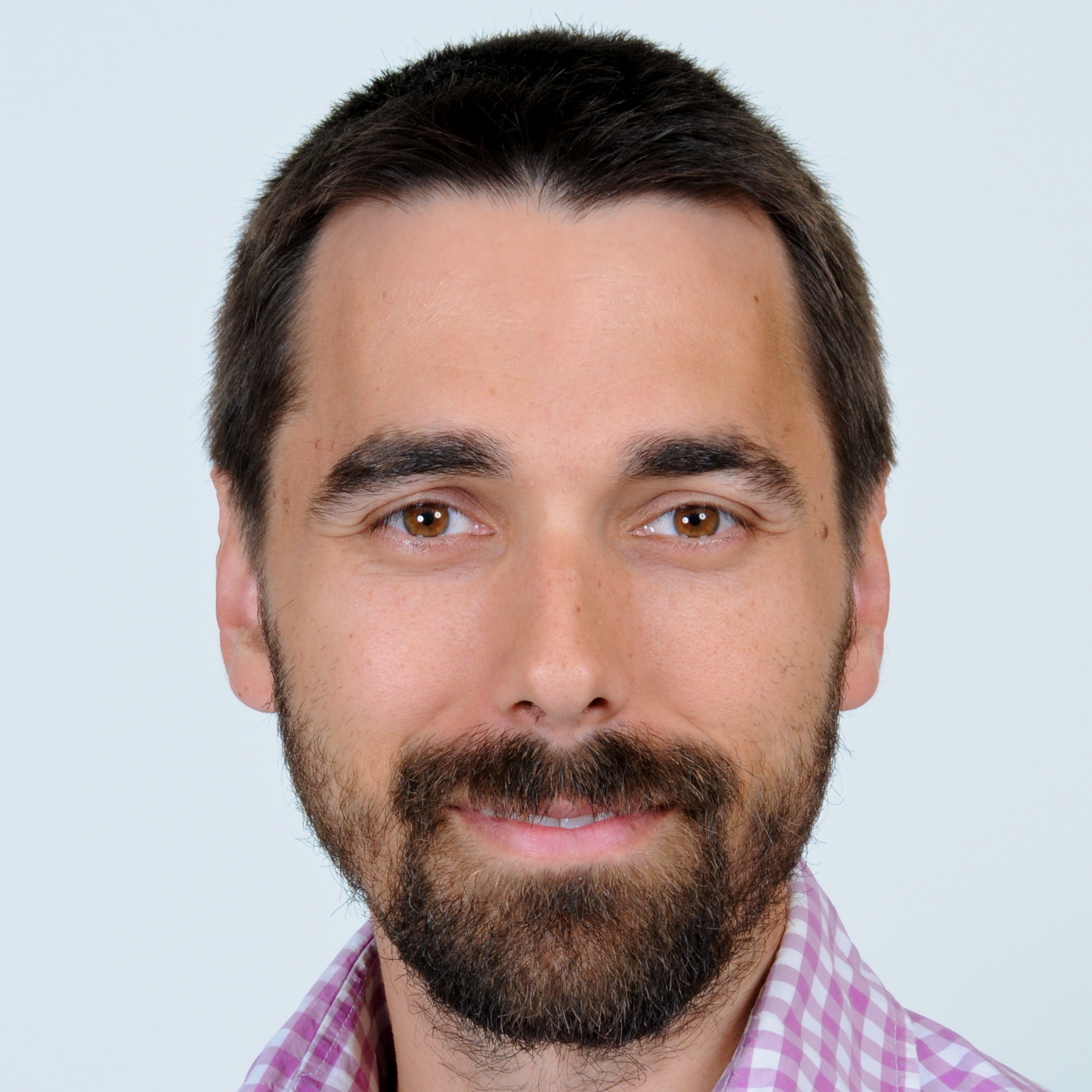}}]{Dr. Justin P. Coon}
received a B.S. degree (with distinction) in electrical engineering from the Calhoun Honours College, Clemson University, USA and a Ph.D. in communications from the University of Bristol, UK in 2000 and 2005, respectively. He has worked in research positions in industry and academia, and is currently an Associate Professor with the Department of Engineering Science, Oxford University, and a Tutorial Fellow of Oriel College. Dr. Coon is the recipient of Toshiba's Distinguished Research Award and two "best paper" awards. He has served as an Editor for the IEEE Transactions on Wireless Communications (2007 - 2013) and the IEEE Transactions on Vehicular Technology (2013 - present). Dr. Coon's research interests include communication theory and network theory.
\end{IEEEbiography}
\begin{IEEEbiography}
    [{\includegraphics[width=1in,height=1.25in,clip,keepaspectratio]{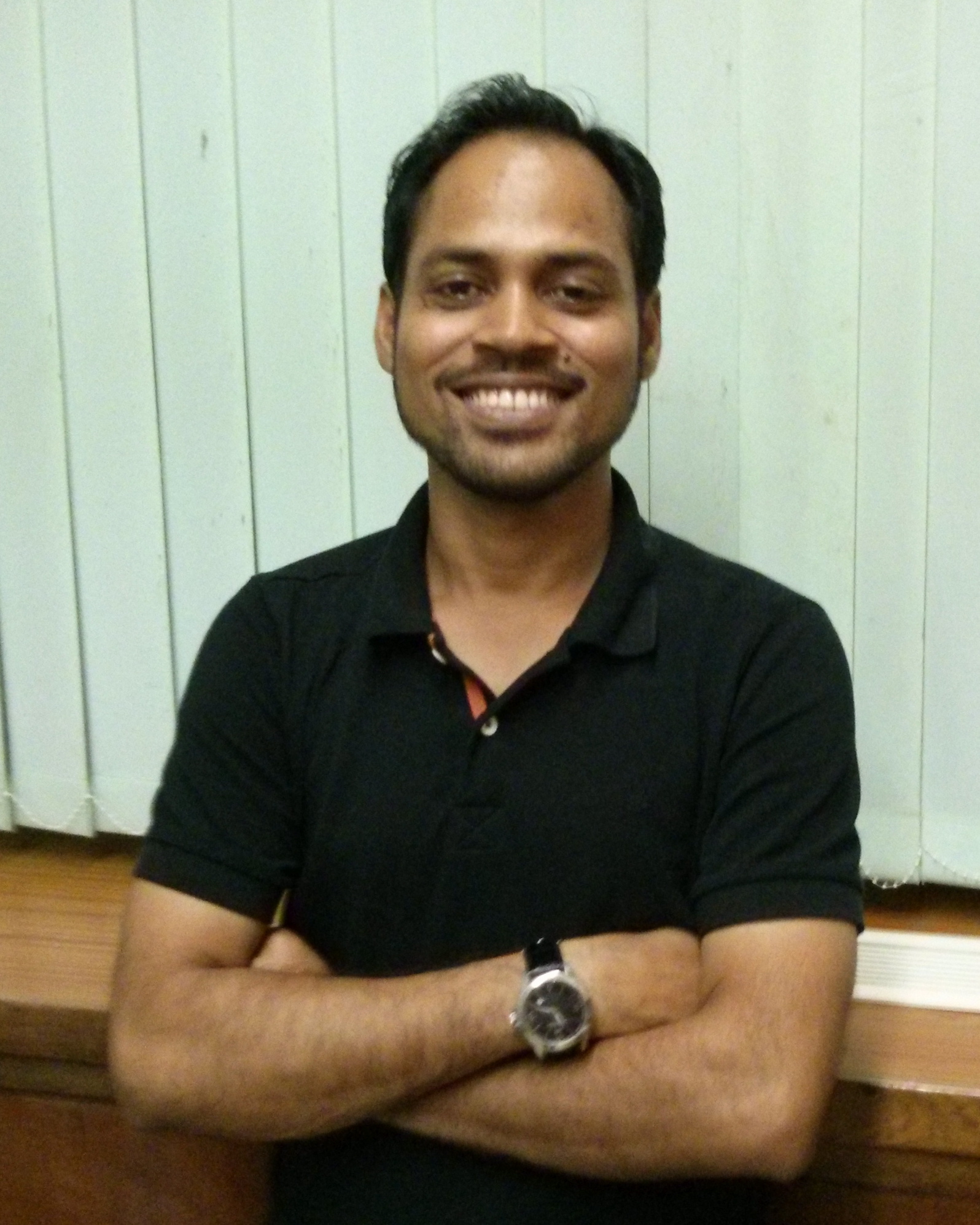}}]{Dr. Naqueeb Warsi} graduated from the Faculty of Engineering and Technology, Jamia Millia Islamia, New Delhi with a B.Tech in Electronics and Communication Engineering in 2006. After that, he worked as a scientist for the Space Application Center (Indian Space Research Organization) until 2009. He then went on to obtain a PhD in information theory from the Tata Institute of Fundamental Research in Mumbai in 2015. Currently he is working as a Postdoctoral Researcher in the Department of Engineering Science at the University of Oxford. Dr. Warsi's research interests lie in the area of classical and quantum information theory, particularly information theoretic problems in the non-asymptotic regime.
\end{IEEEbiography}

\end{document}